\documentclass{emulateapj}
\usepackage{graphicx}
\usepackage{subfigure}
\usepackage{epsfig}
\usepackage{times}
\usepackage{natbib}
\usepackage{amsfonts}
\usepackage{amsmath}
\usepackage{amsbsy}
\usepackage{hyperref}
\usepackage[stable]{footmisc}
\usepackage{color}
\usepackage{booktabs}
\begin{document}
\newcommand{\project}[1]{\textsl{#1}}
\newcommand{\fermi}{\project{Fermi}}
\newcommand{\rxte}{\project{RXTE}}
\newcommand{\rhessi}{\project{RHESSI}}
\newcommand{\hz}{\,\mathrm{Hz}}
\newcommand{\com}[1]{\noindent\textcolor{blue}{#1}}
\newcommand{\prop}[1]{\noindent\textcolor{red}{#1}}

\title{The accretion rate dependence of burst oscillation amplitude}
\author{Laura S. Ootes\altaffilmark{1}, 
Anna L. Watts\altaffilmark{1}, 
Duncan K. Galloway\altaffilmark{2}, 
Rudy Wijnands\altaffilmark{1}
}
\affil{\altaffilmark{1} Anton Pannekoek Institute for Astronomy, University of Amsterdam, Postbus 94249, 1090 GE Amsterdam, The Netherlands\\
\altaffilmark{2} Monash Centre for Astrophysics (MoCA) and School of Physics, Monash University, Clayton, Victoria 3800, Australia}
\email{l.s.ootes@uva.nl}

\begin{abstract}
\noindent Neutron stars in low mass X-ray binaries exhibit oscillations during thermonuclear bursts, attributed to asymmetric brightness patterns on the burning surfaces. All models that have been proposed to explain the origin of these asymmetries (spreading hotspots, surface waves, and cooling wakes) depend on the accretion rate. By analysis of archival \textit{RXTE} data of six oscillation sources, we investigate the accretion rate dependence of the amplitude of burst oscillations. This more than doubles the size of the sample analysed previously by \citet{muno2004}, who found indications for a relationship between accretion rate and oscillation amplitudes. We find that burst oscillation signals can be detected at all observed accretion rates. Moreover, oscillations at low accretion rates are found to have relatively small amplitudes ($A_\text{rms}\leq0.10$) while oscillations detected in bursts observed at high accretion rates cover a broad spread in amplitudes ($0.05\leq A_\text{rms}\leq0.20$). In this paper we present the results of our analysis and discuss these in the light of current burst oscillation models. Additionally, we investigate the bursts of two sources without previously detected oscillations. Despite that these sources have been observed at accretion rates where burst oscillations might be expected, we find their behaviour to be not anomalous compared to oscillation sources. 

\end{abstract} 

\section{Introduction}
\label{sec:introduction}
\noindent Neutron stars in low mass X-ray binaries (LMXBs) accrete matter from their companion via Roche-lobe overflow. As the hydrogen- and helium rich matter accumulates on the surface of the neutron star, it is compressed. For accretion rates $10^{-3}\dot{m}_\text{Edd}\leq \dot{m}\leq 1 \dot{ m}_\text{Edd}$ (where $\dot{m}_\text{Edd}=8.8\times10^4\text{ g cm}^{-2}\text{ s}^{-1}$ is the local Eddington accretion rate), the pressure increase leads to a thermonuclear instability which initiates unstable ignition of the accreted material \citep[see reviews by][]{bildsten1998,galloway2008}. This results in a runaway event known as a type I X-ray burst, during which the material in the accretion layer is fused into heavier elements. To date, over 100 type I X-ray burst sources have been observed.\footnote{On-line catalogue of type I X-ray burst sources by J. in 't Zand (SRON): \url{www.sron.nl/~jeanz/bursterlist.html}}

One of the first discoveries obtained from observations with the \textit{Rossi X-ray Timing Explorer} (\textit{RXTE}) was that of thermonuclear burst oscillations \citep{strohmayer1996}. These are periodic fluctuations in luminosity at a close to stable frequency that can arise during any phase of a type I X-ray burst. Burst oscillations have been observed in 18 sources to date, and for these sources the phenomenon is not necessarily detected in each burst.\footnote{Burst oscillation library by A. L. Watts (UvA): \url{https://staff.fnwi.uva.nl/a.l.watts/bosc/bosc.html}} The fractional rms amplitude of the signal typically has a value in the range $0.05\leq A_\text{rms}\leq0.20$. A characteristic property of burst oscillations is that the frequency of the signal tends to drift smoothly upwards by $1-3$ Hz during the course of a burst, towards an asymptotic maximum that is nearly constant for each source \citep{muno2002a}. The discovery of burst oscillations in the 401-Hz accretion powered pulsar SAX J1808.4-3658 \citep{chakrabarty2003} revealed that the oscillation frequency is close to the spin frequency of the pulsar in that system \citep{wijnands1998}. Although individual spin measurements have not been obtained for all oscillation sources, it is generally assumed that the oscillation frequency represents the spin frequency of the neutron star. The discovery of oscillations in SAX J1808.4-3658 confirmed the proposed theory that thermonuclear burst oscillations are caused by asymmetric brightness patterns on the burning surface of a neutron star during a type I X-ray burst. 

The origin of the brightness asymmetries is an open question. Various models have been proposed that try to explain the underlying mechanism, and these can be divided into three main (non-exclusive) categories: hotspot models, surface wave models, and cooling wake models \citep[see review articles by][]{strohmayer2006,watts2012}. Two factors on which the presence of a surface asymmetry depends are the ignition location of the burst, and the way that the flame spreads. It is generally assumed that the burst is ignited at one point on the surface of the neutron star, since the accretion time between two bursts is much longer than the time required for a thermonuclear instability to develop \citep{shara1982}. This makes it unlikely that the thermonuclear instability that initiates the burst arises everywhere on the surface at the same time, considering the level of thermal homogeneity that would otherwise be required. \citet{spitkovsky2002} showed that the ignition occurs preferentially at the equator of the star rather than at higher latitudes, because of the reduced effective gravity force at this location.  However, for specific accretion rates, or in cases of strong magnetic channeling, the preferred ignition latitude ($\phi_\text{ign}$, where $\phi=0$ indicates the equator) is predicted be off-equatorial \citep{cooper2007, maurer2008} or even near the magnetic poles \citep{cavecchi2016}.  How the flame subsequently spreads determines for how long an asymmetry can persist. An important question is whether the flame spread covers the whole surface or is confined to a smaller region. The flame spread depends both on the heat transfer mechanisms and the hydrodynamical effects involved.  The two main factors that influence both the longitudinal and latitudinal propagation are the conductivity and Coriolis force  \citep{spitkovsky2002,cavecchi2013,cavecchi2015}.   Additionally, \citet{cavecchi2016} showed that magnetic fields can significantly affect flame propagation.

In the hotspot models, it is assumed that the burst starts at one point on the surface of the neutron star (most likely at the equator) after which the flame spreads in all directions. If the hotspot arises on or close to the equator rather than at one of the rotational poles, the hotter region forms an azimuthal asymmetry and will be observed as an oscillation with a frequency close to the spin frequency of the neutron star. The growing hotspot can either engulf the entire star, after which the asymmetry is resolved, or it can be confined to a small region on the surface by various mechanisms \citep[see][and references therein]{watts2012} such as Coriolis force confinement \citep{spitkovsky2002,cavecchi2013,cavecchi2015} or magnetic confinement \citep{cavecchi2016}\footnote{Magnetic confinement is certainly implicated in IGR J17480-2446, a system with burst oscillations that rotates too slowly for  Coriolis confinement to be effective \citep{cavecchi2011}.}.  While spreading/confined hotspot models are supported by various aspects of the observations \citep{galloway2008,watts2008,cavecchi2011,chakraborty2014}, they cannot easily explain oscillations far in the tails of X-ray bursts, the absence of detected oscillations in some bursts, and the largest observed frequency drifts. 

Surface wave models assume that the large scale waves are excited in the outer layers of the neutron star (global modes) \citep{heyl2004}. These cause height differences in the burning layers, which can be observed as brightness patches. The waves are excited as soon as the initial hotspot starts to spread, and can persist after the flame has engulfed the entire surface of the star. The main difficulty with these models is that the predicted frequency drifts (which occur naturally as the surface layers cool) are too large compared to the observations \citep{piro2005a,piro2005b,berkhout2008}.  In addition self-consistent models of mode excitation to sufficient amplitude are still required \citep[although see][]{narayan2007}.

In the cooling wake models, it is assumed that burst oscillations are caused by sequential cooling of different regions (for example, with the regions that ignite first cooling first).  If the cooling timescale is independent of position this cannot produce oscillations of sufficient amplitude:  some degree of asymmetric (position-dependent) cooling would be required to reproduce the observed amplitudes \citep{cumming2000,mahmoodifar2016}.  Physical mechanisms that might lead to asymmetric cooling include transverse heat flows, or variations in the column depth at which the heat is released, which depends on the local depth of the accreted layer.   An alternative, suggested by \citet{spitkovsky2002}, is that cooling wakes might drive zonal flows, which then induce atmospheric vortices \citep[see also the discussion in][]{zhang2013}.  Note that cooling wake models cannot, of course, explain oscillations in the rising phase of bursts.  

The proposed burst oscillation mechanisms depend at least in part on the (local) mass accretion rate onto the neutron star. The presence of unstable burning regimes, set by the local accretion rate, determine the ignition latitude (for example, whether ignition occurs on or off the equator), and hence whether or not the initial hotspot causes a brightness asymmetry. Ignition latitude in turn is an important factor in the various flame confinement models. For surface wave models, the accretion rate may be related to the question or whether or not the mode is unstable enough to grow to significant size \citep{narayan2007}. Additionally, the accretion rate might be an important factor in the asymmetric cooling mechanism. \citet{muno2004} showed that the detectability of burst oscillations is not determined by the properties of the X-ray bursts they occur in. Instead, they found that the oscillations seemed to occur preferentially when the source is in a higher accretion state.  Additionally, they found that the upper limits on the amplitude of the oscillations appeared to be larger at high accretion rate, which led them to suggest that the amplitudes are attenuated at low accretion rate. However, the amount of data that was available at the time was insufficient to constrain the exact relationship between the two parameters.

To explain why the observed amplitudes seemed to be smaller at low accretion rate, \citet{muno2004} considered whether the presence of an electron corona with optical depth $\tau\approx3$ at low accretion rate could be the origin of the attenuation of the oscillations. Such a corona could scatter the photons from the neutron star surface, reducing the amplitude by a factor of two. However, as they pointed out, the electron corona is expected to be even more optically thick at higher accretion rates. Therefore, their suggestion requires that the geometrical configuration of the corona at high accretion rate is such that it prevents photon scattering. Since there are no indications how such a change in configuration would be possible, this extra condition makes the suggested cause of oscillation attenuation rather unlikely. 

In this research we investigate the type I X-ray bursts of six different burst oscillation sources observed with \textit{RXTE}, with the goal of constraining the relationship between burst oscillation amplitude and accretion rate. We investigate the same sources as \citet{muno2004}, but extend the burst sample significantly compared to this research from 333 bursts to 765, such that for most sources more than twice the amount of bursts is analysed. We compare our results to the expectations of various burst oscillation models, to gain insight on the thermonuclear burst oscillation mechanism. Additionally, we investigate the bursts of two LMXBs for which type I X-ray bursts have been observed over a wide spread of accretion rates, but for which to date no burst oscillations have been detected. \citet{watts2012} raised the question whether or not these two sources might be anomalous in their behaviour compared to the oscillation sources, because bursts from these two sources have been observed at an accretion rate limit above which most detections are found \citep{galloway2008}. For these two sources, we carry out a similar analysis compared to the oscillation sources in order to determine upper limits on the oscillation amplitudes with the goal of understanding why no oscillations have been detected in these sources so far.  

\section{Observations}
\label{sec:data}

\subsection{Telescope and on-line data catalogues}
The burst sample that we analyse consists exclusively of observations from \textit{RXTE}. Type I X-ray bursts are detected with the Proportional Counter Array (PCA). The PCA consists of five xenon-filled proportional counters, which are sensitive to photons with energies in the range $2-60$ keV \citep{jahoda1996}. 

In this research we made use of information derived from bursts collected in the on-line \textit{RXTE} catalogue \citep{galloway2008} and the Multi-INstrument Burst ARchive (MINBAR), such as burst start times, spectral state of the source during the observation, and notes of any peculiarities of the bursts or observations. The MINBAR database contains information on all X-ray bursts observed with \textit{RXTE}, \textit{BeppoSAX}, and \textit{INTEGRAL}.\footnote{The MINBAR database, maintained by Dr. D. Galloway, can be found at \url{http://burst.sci.monash.edu/minbar}.} We collected the event mode \textit{RXTE} data from the public NASA archive.\footnote{ \textit{RXTE} public data archive: \url{ftp://legacy.gsfc.nasa.gov/FTP/xte/data/archive/}}

\subsection{Sources}
We investigate the bursts of eight LMXBs. Six of these sources are confirmed burst oscillation sources for which some part of the data presented in the current paper was previously analysed by \citet{muno2004}: 4U 1608-52, 4U 1636-536, 4U 1702-429, 4U 1728-34, KS 1731-26, and Aql X-1. The two other sources that we investigate are 4U 1705-44 and 4U 1746-37. These sources exhibit type I X-ray bursts, but have not been observed to show burst oscillations. 

All sources are Galactic LMXBs, but the distances to these sources are not well constrained. \citet{galloway2008} estimated the distances to 48 LMXBs, including the eight in our sample, from their PRE bursts, but since this requires knowledge of the mass, radius, and in particular the composition of the atmosphere of the neutron stars, the resulting distance ranges are rather large.  Based on their spectral behaviour, all sources are classified as atoll sources \citep{hasinger1989}, tracing out a wide range of accretion rates.  Three of the investigated sources are transient sources, the remaining five sources are persistent X-ray emitters. Aql X-1 differs from the other sources, because this source is the only intermittent X-ray pulsar in the sample. The source has shown one rare incidence of intermittent accretion-powered pulsations \citep{casella2008}. 

\subsection{Burst sample}\label{sample}
In total we investigated 889 bursts from \textit{RXTE} in this research. Initially, we obtained all available burst data of the selected sources from the public \textit{RXTE} archive, in order to cover the largest possible accretion rate range and to obtain the highest possible statistical significance on any observable trend in the data. Subsequently, we discarded bursts from the sample based on the following criteria:
\begin{itemize}
\item{We eliminated all bursts that are marked with one of the following flags in either the \textit{RXTE} or MINBAR database: e, f, g, h \citep{galloway2008}. These flags indicate: e) Very faint bursts, for which only the burst peak could be observed, and no other parameters could be determined. f) Bursts that are either very faint or bursts for which there were problems with the background subtractions,  such that no spectral fit of the burst could be obtained. g) Bursts that we only partly observed, resulting in an unconfirmed burst. h) Bursts that were not covered by the high time resolution data modes of the telescope. A total of 36 bursts was eliminated from the sample based on these flags.}
\item{We set a minimum background-subtracted burst count of 5000 counts within the first 16 seconds of the burst. This limit ensures that each burst can be divided in at least one full time bin (see Section \ref{binning}). We excluded 57 bursts that did not meet this criterion from further analysis.}
\item{We determined bursts with gaps in the data that lasted for multiple seconds to be unfit for analysis. The \textit{RXTE} catalogue does provide a flag that indicates that the burst contains data gaps. However, we did not eliminate all bursts with this flag, but only those where the burst gap is so large that it affects the outcome of the burst analysis, which is the case for data gaps $\gtrsim 1$ sec. The main problem with bursts with such large data gaps is that the gaps eliminate one or more full time bins (as defined in section \ref{binning}) from the burst. This means that there is a significant chance that the time bin with the strongest signal is lacking from the burst, which would affect the outcome of the analysis. We eliminated seven bursts from the sample based on this criterion.}

\item{Bursts that are not fully observed by \textit{RXTE} were eliminated from the sample. These coincide with the bursts with label g in the \textit{RXTE} and MINBAR database. However, there are bursts in the sample without this label for which (part of) the last phase before the start of the burst or the burst decay were not observed. Since we perform our analysis based on the 17 seconds before the start of the burst (to determine the background count rate) up to 16 seconds after, we eliminated three partially observed bursts that were not flagged with the g label by hand to ensure that all bursts are analysed in a homogeneous way.  }
\end{itemize}

\noindent The properties of the burst sample are displayed in Table \ref{bursts}, including the number of bursts eliminated from the sample for each source. We list the number of bursts analysed by \citet{muno2004} for comparison to show that we do indeed significantly extend the number of analysed bursts. No new bursts are available for KS 1731-26, since \textit{RXTE} observations showed that the source returned to quiescent state in February 2001 \citep{wijnands2001}. We analyse for this source the same bursts as in the previous research. This way we can observe what kind of influences the small changes in analysis method have on the results. Table \ref{bursts} also provides 
the frequency of the detected oscillations. 

\begin{deluxetable}{lcccc}
\tabletypesize{\small}
\tablewidth{\columnwidth}
\tablecaption{Burst sample}
\tablehead{\colhead{Source} & \colhead{$N_\text{bursts}$ prev.$^{1}$} & \colhead{$N_\text{bursts}$} & \colhead{Eliminated} & \colhead{$\nu_\text{o}\text{ }^1$}\\
\colhead{} & \colhead{} & \colhead{} & \colhead{} & \colhead{(Hz)}}
\startdata
4U 1608-52  &  28  &  56 & 9  &  620  \\
4U 1636-636  &  124  &  381  & 42  &  581  \\
4U 1702-429  &  18  &  50 & 1  &  329  \\
4U 1705-44  &  -  &  94 & 24  &  -  \\
4U 1728-34  &  104  &  176 & 15  &  363 \\
KS 1731-26  &  27  &  27  & 0 &  524  \\
4U 1746-37  &  -  &  30 & 8  &  -  \\
Aql X-1  &  32  &  75 & 4  &  549
\enddata
\tablecomments{Apart from the number of bursts in the initial sample of this research ($N_\text{bursts}$), the number of bursts analysed by \citet{muno2004} is displayed as well ($N_\text{bursts}$ prev.: all bursts observed by \textit{RXTE} up to 2003 August). The latter is displayed for comparison, to stress that we significantly extend the amount of analysed bursts.} 
\tablerefs{$^1$ \citet{muno2004}}
\label{bursts}
\end{deluxetable}

\section{Method}% } 
\label{sec:method}

\subsection{Accretion rate}\label{sz}
The local mass accretion rate ($\dot m$) of accreting sources can be estimated from the persistent (between bursts) X-ray luminosity ($L_\text{x}$) of the source: 
\begin{equation}
L_\text{x}=\frac{4\pi R^2\dot{m}Q_\text{grav}}{(1+z)}\end{equation}\label{accretionrate}
with $R$ the radius of the neutron star, $Q_\text{grav}$ the energy released per nucleon during accretion and $z$ the surface redshift \citep[see][]{galloway2008}. 

\begin{figure}[t!]
	\begin{center}
	\includegraphics[width=\columnwidth]{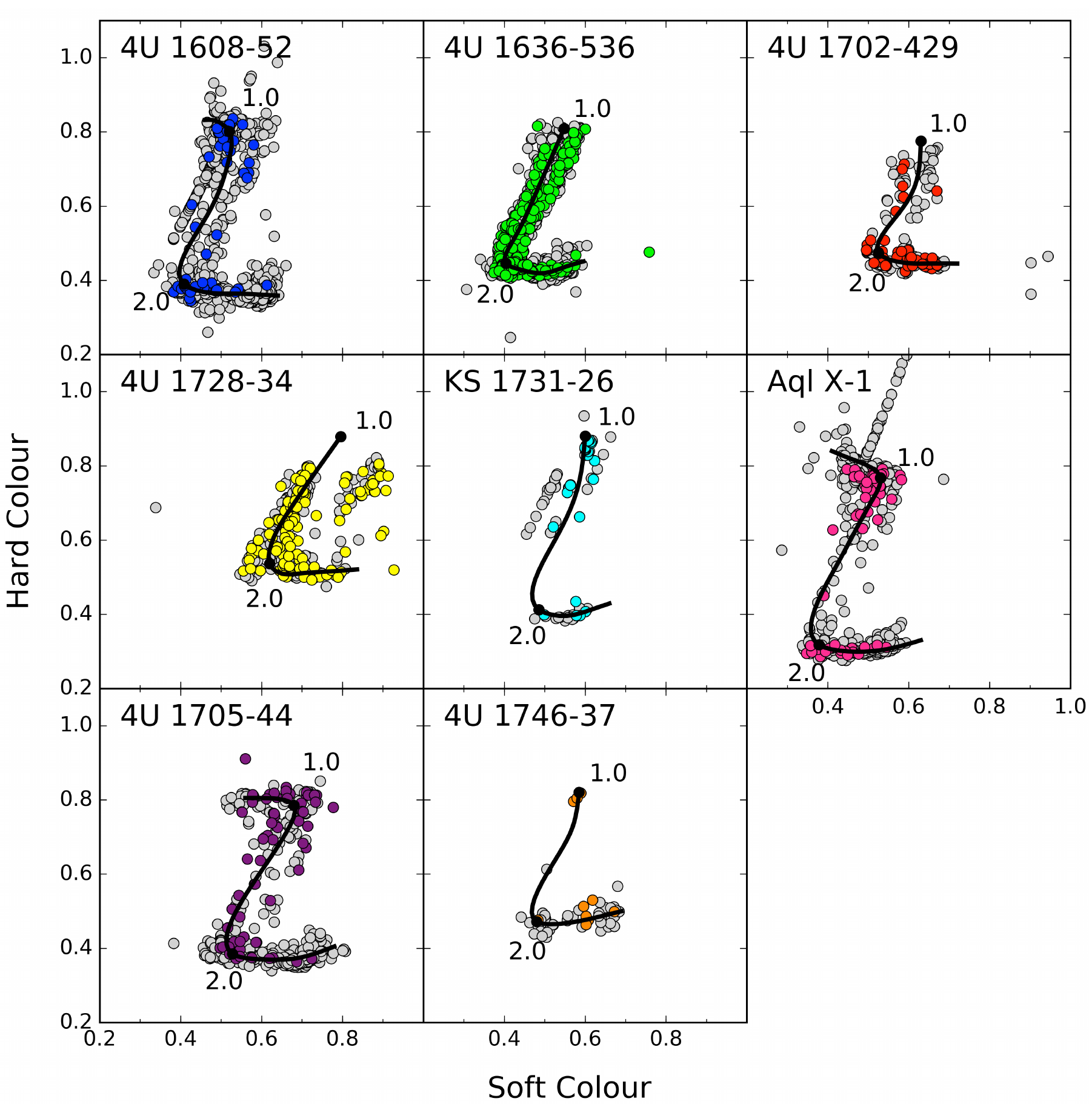}
	\end{center}
	\caption{Colour-colour diagrams for each of the eight analysed sources, taken from MINBAR. Each point represents the spectral state of the source during a \textit{RXTE} observation. Coloured dots are observations with one or multiple bursts, grey dots are observations without detected bursts. The $S_\text{Z}$ values (a measure of accretion rate) of the bursts are derived from their position relative to the $S_\text{Z}$-curve (black curve). By definition, the upper right corner of the $S_\text{Z}$-curve is assigned the value $S_\text{Z}=1.0$ and the lower left corner of the Z-shaped path $S_\text{Z}=2.0$ as indicated in the figure.  }\label{colours}
\end{figure}

For atoll sources, it is thought that a measure of accretion rate can be obtained from the position of the source in its colour-colour diagram \citep{hasinger1989}. This measure is determined by the spectral state of the source, the emission in hard X-rays (high energy) versus soft X-rays (low energy). Figure \ref{colours} shows the colour-colour diagrams from the eight investigated sources, using colour data from the MINBAR database. The colours from the MINBAR database are determined from the spectral model: the ratios of integrated flux (based on the best-fit spectral model) are calculated in different energy bands. This is different from other approaches \citep[e.g.][]{galloway2008} that calculate the ratios of counts  in different energy bands rather than integrated flux. While the method used for MINBAR has a dependency on the spectral model, it has the advantage is that it is independent of the instrument used to obtain the data.

Atoll sources move along a specific path in a colour-colour diagram while the persistent luminosity increases \citep[e.g.][]{homan2010}. The spectral state of an atoll source is indicated by the value $S_\text{Z}$. This value is obtained from the parametrisation of the path that the source traces out \citep[see][for details of this method]{mendez1999, galloway2008}. By definition, the upper right corner of the path is assigned $S_\text{Z}=1$ and the lower left corner $S_\text{Z}=2$ (see the colour-colour diagrams that were taken from the MINBAR database shown in Figure \ref{colours}, in which we included for each source the $S_\text{Z}$-curve with the assigned corner points). The $S_\text{Z}$ values of all other points in the diagram are extrapolated from these two. High $S_\text{Z}$ corresponds to high accretion rate. In this analysis we use $S_\text{Z}$ values from the MINBAR database and \textit{RXTE} catalogue. 

Although it is still under debate whether or not $S_\text{Z}$ is the best measure of mass accretion rate, it has the advantage that is does not require estimates of the distance to the source, and the mass and radius of the neutron star.  Some caution is required when comparing results from different stars since it might not necessarily be that a given $S_\text{Z}$ value for one source corresponds to the same accretion rate as for another.  Spectral state was also used  as a measure of accretion rate in the previous analysis by \citet{muno2004}, allowing us to determine the influence of a larger sample size on those results. 

\subsection{Data analysis}

We analyse each burst of the oscillation sources individually to determine whether an oscillation can be detected. We look for signals in the first 16 seconds of the burst with a frequency within 5 Hz of the known oscillation frequency ($\nu_o\pm 5$ Hz) to account for any frequency drift. Although in most cases the frequency drift is only $1-3$ Hz \citep{muno2002a}, larger drifts have been reported as well \citep{rudy2001}. We use the range $\nu_\text{o}\pm5$ Hz to be able to detect the largest possible drifts and to be consistent with the expected drifts from the proposed oscillation models.  In case of a detection (see Section \ref{detectcrit}), the fractional root mean square amplitude (rms amplitude) of the signal is computed. For those bursts in which we do not detect an oscillation signal that passes the detection criterion, we compute an upper limit on the rms amplitude.

In the following subsections the analysis method of bursts from the oscillation sources is described sequentially. The same method is applied to the sources without detected oscillations, but with a few adjustments since the spin frequency of these sources is unknown. The exact treatment of the bursts from these two sources is described in Section \ref{newsources}. 

\subsubsection{Burst start time and background count rate}\label{bst and bg}
First we compute for each burst the burst start time ($t_0$) and the background count rate. We estimate the background count rate using the count rate in the range $20 - 5$ seconds prior to the approximate burst start time given in one of the databases. $t_0$ is then defined as the time where the count rate equals 1.5 times the estimated background count rate.  This ensures that all the burst start times are defined by the same criterion.  Next, the true background count rate ($C_\text{B}$) is defined as the average count rate in the range $17-1$ seconds preceding $t_0$. \citet{muno2004} used the 16 seconds directly prior to the burst to compute the background. A time buffer of one second is kept between the burst start time and the range from which the background is calculated to ensure that the background is not overestimated in bursts with a slow rise.

\subsubsection{Binning}\label{binning}
Secondly, the first 16 seconds of the burst, $[t_0-(t_0+16.0\text{ sec})]$, is divided into non-overlapping time bins with 5000 counts each. We divided the bursts into time bins with equal amount of counts to make the error bars on each measurement similar \citep{watts2005}.  Note that in a previous analysis by \citet{muno2004} equal time bins were used, so that error bars later in the burst were larger (see discussion in Section \ref{methodeffects}).   The number of time bins in our analysis thus depends on the strength of the burst and the underlying background. We use non-overlapping time bins to ensure that each time bin is independent of the others. This makes it easier to compute the number of trials to obtain a signal (see Section \ref{detectcrit}).

In each time bin we look for signals within 5 Hz of the known oscillation frequency. For each time bin we set up 10 frequency bins ($\nu_\text{o}\pm5$ Hz), to obtain a frequency resolution of 1 Hz (equal to the resolution in \citet{muno2004}). We thus create for each burst a two-dimensional grid of time-frequency bins in which we attempt to detect oscillation signals (see Figure \ref{grid} for a visualisation of the grid).

\subsubsection{Measured power}
We compute for each time bin the signal power for each of the 10 trial frequencies. We obtain the measured power for a signal with trial frequency $\nu$ by calculating the $Z^2$ statistic \citep[see][]{buccheri1983, strohmayer1999}. The $Z^2$ statistic is similar to a fast Fourier transform in the sense that it decomposed a signal into the sines and cosines that it consists of (see Equation \ref{zsquare}). This results in a power spectrum in which the power of the signal is plotted as function of frequency. The difference with a fast Fourier transform is that $Z^2$ statistics does not require that the arrival times of the counts are binned. 
The $Z^2$ statistic is defined as:

\begin{equation}\label{zsquare}
Z^2_n=\frac{2}{N_\gamma}\sum_{k=1}^n\left[\left(\sum_{j=1}^{N_\gamma} \cos{k\nu t_j}\right)^2+\left(\sum_{j=1}^{N_\gamma} \sin{k\nu t_j}\right)^2\right]
\end{equation}
where $Z^2$ is the measured power of the signal, $n$ is the number of harmonics, $N_\gamma$ is the number of counts in the time bin, and $t_j$ the arrival time of the j-th count relative to some reference time. We only look for the first harmonic of each signal, so $n=1$. By definition of the time bins $N_\gamma=5000$. 

Using this statistic, we obtain a power spectrum for each time bin in which the power of the oscillation signals is plotted as function of the 10 trial frequencies. From such a spectrum one can easily determine at which frequency the signal is strongest. However, when frequency drifts between Fourier bins, the computed amplitude drops artificially \citep{vanderklis1989}. We assume that the frequency of the oscillation is constant within each time bin and thus do not take into account frequency drifts within each bin. 

\subsubsection{Detection criteria}\label{detectcrit}

\begin{figure}
	\begin{center}
	\includegraphics[width=\columnwidth]{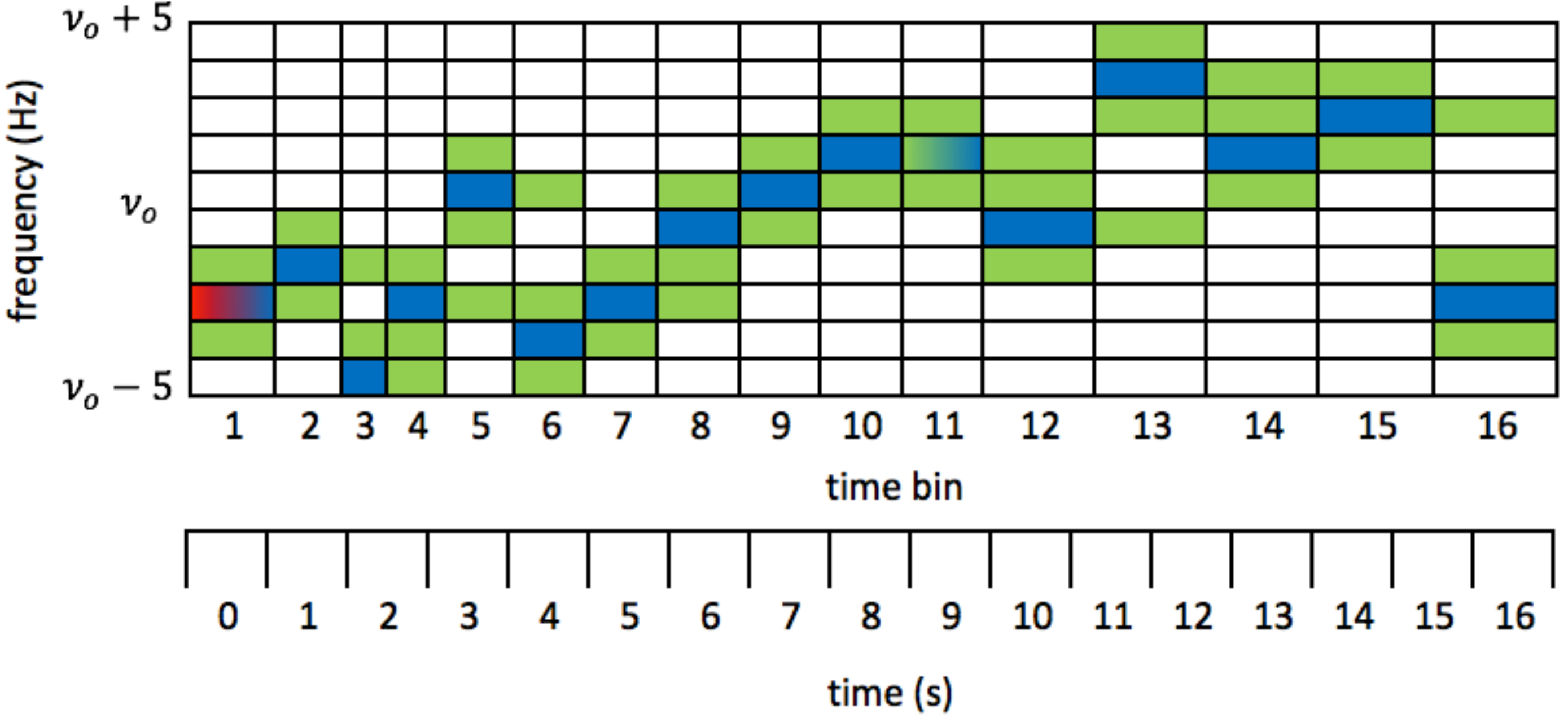}
	\end{center}
	\caption{Visualisation of the time-frequency grid created to search for a burst oscillation. The first 16 seconds of a burst are divided into time bins with 5000 counts each. This means that the time bins are not equally broad in time space. For each time bin there are 10 frequency bins ranging from $\nu_o-5.0$ Hz to $\nu_o+5.0$ Hz, such that we look for signals within 5 Hz of the known oscillation frequency ($\nu_o$). The colours indicate which selection criterion applies to that bin. Blue = criterion 1: single bin detection (largest measured power of that time bin). Red = criterion 2: first bin detection. Green = criterion 3: double bin detections. Each green bin is an adjacent time or frequency bin to a blue bin, with which it forms a pair of double bins. Each blue bin has three adjacent bins which we use to look for double bin signals: both of the adjacent frequency bins and the next time bin.}\label{grid}
\end{figure}

For each individual time bin of a burst, we select the frequency bin with the largest measured power and determine whether or not the signal is considered a detection.  We assume a Poisson noise process, for which powers in the absence of a signal are distributed as $\chi^2$ with two degrees of freedom.  This assumption is reasonable at high frequencies, but not at low frequencies where the red noise contribution due to the burst light curve envelope becomes significant. For this reason, we do not look for oscillation signals below 50 Hz in the two sources without previously detected oscillations.  Based on the assumption for noise distribution, we can then determine the chance that any measured power is produced by noise alone. We can then set a threshold for the measured power above which we define a signal to be significant. We choose to set the detection criterion such that the chance that a signal was produced by noise is less than 1\% when taking into account the number of trials for each burst $(N)$. The number of trials in defined as the total number of time-frequency bins in which one looks for a signal; where $N=N_\text{t}*N_\nu$ with $N_\text{t}$ the number of time bins and $N_\nu$ the number of frequency bins. 

The probability (Prob) that a measured signal with noise chance $\delta$ was produced by noise for $N$ trials is given by:
\begin{equation}
\text{Prob}=N\delta(1-\delta)^{N-1}
\label{prob}
\end{equation}
Based on the detection criterion for each burst, we can define three criteria, similar to \citet{muno2004}, for which we determine a measured power to be a significant detection (see Figure \ref{grid} for a visualisation of each criterion):
\begin{enumerate}
\item The chance that a measured power $Z_m$ was produced by noise is less than $7\times10^{-5}$ in a single trial ($\delta\leq 7\times 10^{-5}$), assuming that a burst will on average consist of 16 individual time bins, such that $N=16*10$. This corresponds for 1\% chance overall to a measured power criterion $Z_m^2\geq 19.4$.  
\item{A signal occurring in the first second of a burst has a single trial chance probability $\delta\leq10^{-3}$. This probability results in a measured power limit $Z_m^2\geq13.8$. This detection criterion was introduced by \citet{muno2004} as well. At the burst onset, the difference in brightness between burning and non-burning material is largest, and therefore oscillation signals would be expected to be largest in the burst rise (first second). }

\item{A signal distributed over two adjacent time-frequency bins has a combined single trial noise chance probability $\delta_1*\delta_2\leq1.3\times10^{-6}$. We check this using the fact that this is similar to a measured power limit of the averaged signal in these two adjacent bins of $\bar{Z}_m^2\geq13.8$. There is a significant chance that a signal does not peak exactly in one time-frequency bin, but is spread over multiple bins instead. Therefore, we select in each time bin the signal with the largest measured power and compute the noise chance of the signal that is spread over the selected time-frequency bin and one of three directly adjacent bins: the same time bin and one of two the adjacent frequency bins, or the same frequency bin and the next time bin.   The chance that both bins consist of noise alone is given by the product of the noise chance probabilities of the two individual bins ($\text{Prob}_{1,2}=\text{Prob}_1(N_1,\delta_1)*\text{Prob}_2(N_2,\delta_2)$). \\
To meet the detection criterion of the burst, the single trial probabilities of the two bins ($\delta_1$ and $\delta_2$) must satisfy the equation for $\text{Prob}_{1,2}$. Using an approximation for $\text{Prob}_{1,2}$ given by Equation \ref{doubleprob} (taking into account that $N_2$ is reduced due to the fact that the second bin has to be selected from one of the three bins surrounding the first bin) yields the solution $\delta_1\delta_2=1.3\times 10^{-6}$ that adjacent bins must satisfy to meet the threshold burst probability $\text{Prob}_{1,2}=10^{-2}$.}
\end{enumerate}

\begin{equation}\text{Prob}_{1,2}\approx 3N_\text{t}^2N_\nu\delta_1\delta_2\label{doubleprob}\end{equation}

Each of the detection criteria satisfies the criterion that, on average, an oscillation signal detected from a single burst has a 1\% chance of being a false detection. If one considers each of the three detection criteria as individual trials, the noise probability would increase to a 3\% chance that a detected oscillation is actually a false detection.

\begin{figure}
	\begin{center}
	\includegraphics[width=\columnwidth]{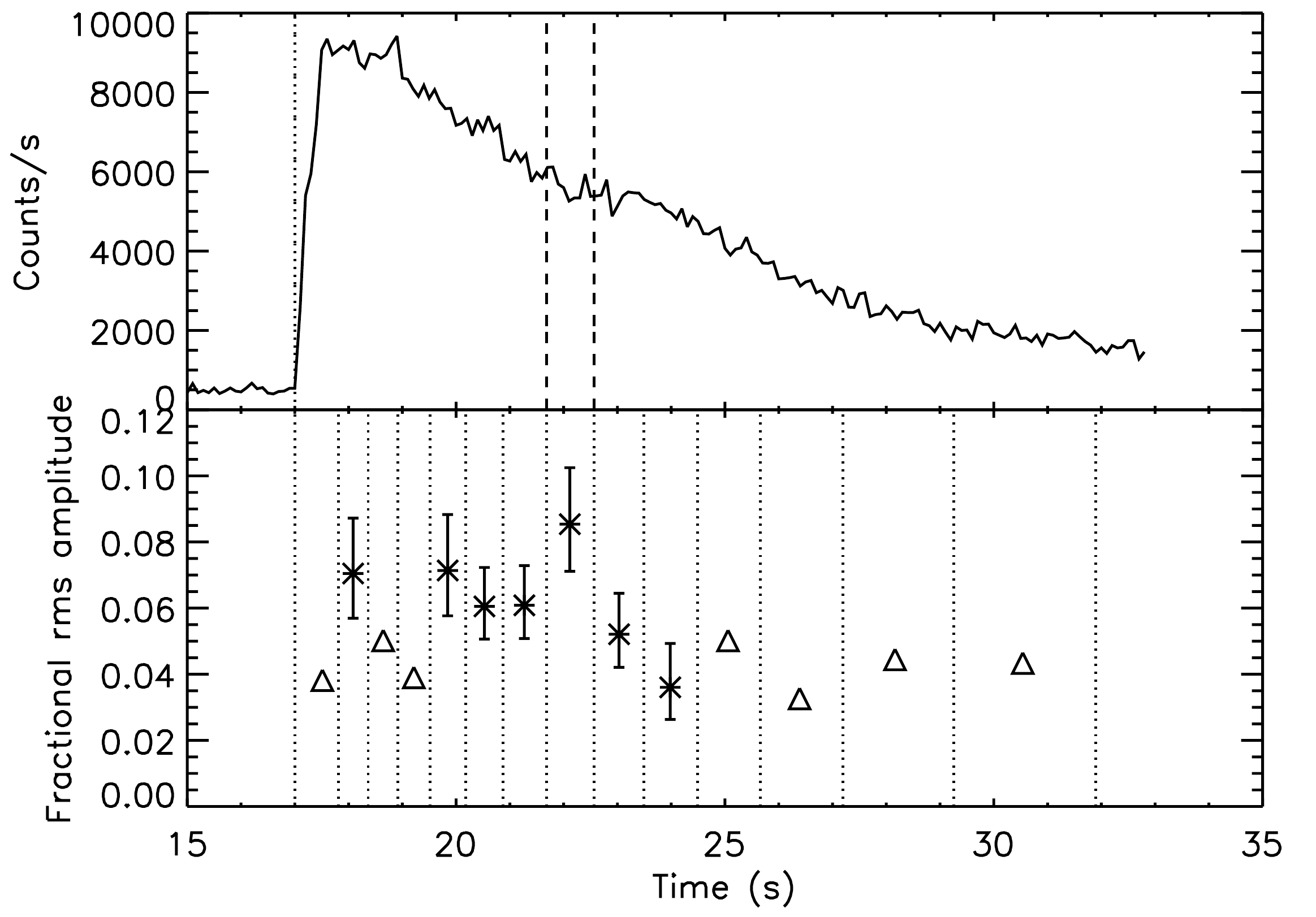}
	\end{center}
	\caption{Result of the analysis of a burst from 4U 1728-34 with observation ID 95337-01-02-00, and $t_0=55474.175$. The upper panel shows the burst itself, and the lower panel shows the limits of the time bins (dotted lines) and in each time bin the computed amplitude (asterisks with vertical error bars) or amplitude upper limit (triangles) in the case of a non-significant signal. In the upper panel the dotted line indicates the burst start time ($t_0$) and the dashed lines represent the time bin in which the oscillation signal with the largest signal power $Z_s^2$ was found.}\label{141}
\end{figure}

\subsubsection{Signal power}
For each time bin we determine four different measured powers, all related to the frequency bin with the largest measured power: we determine the single bin measured power, which can either be in the first second (criterion 2) or at any other timespan in the analysed part of the burst (criterion 1), and three double bin measured powers (criterion 3). However, each measured power consists of two components, the signal power and the noise power. To obtain the signal power and its uncertainty, we correct the measured power for the noise component. We do this using the method outlined in Section 2 of \citet{watts2005} to compute the signal power of a given measured power and the number of harmonics.  

The signal power is derived using the probability distribution $p_n$ of measured signals $Z_m$ for given signal power $Z_s$:
\begin{multline}
p_n(Z_m | Z_s)=\frac{1}{2}\exp{\left[-\frac{(Z_m+Z_s)}{2}\right]} \left(\frac{Z_m}{Z_s}\right)^{(n-1)/2}\times\\
I_{n-1}\left(\sqrt{Z_mZ_s}\right)
\end{multline}
where $n$ is the number of harmonics (we always use $n=1$), and $I$ is a first kind modified Bessel function. The computational procedure provides a signal power and 1-$\sigma$ errors. 

\subsubsection{Oscillation amplitude}\label{oscamp}
The oscillation amplitude of the signal in each time bin is computed from the signal power. As mentioned, there are five possibilities to pass the detection criterion: one from the first criterion, one from the second and three from the third. Per time bin we select from the five options the signal with the largest (averaged) measured power that passed the detection criterion. From the signal power of this oscillation, we compute the fractional rms amplitude of the oscillation ($A_\text{rms}$) using equation \ref{amp}.

\begin{equation}
A_\text{rms}=\sqrt{\frac{Z_s^2}{N_\gamma}}\left(\frac{N_\gamma}{N_\gamma - B}\right)
\label{amp}
\end{equation}

The second term in equation \ref{amp} is the factor that corrects for the background emission, where $N_\gamma$ is the number of counts, and $B$ is the estimated number of background counts in the investigated time bin ($N_\gamma=5000$ and $B=C_B*\Delta t$, with $\Delta t$ the time width of the bin(s) over which the signal is considered). We calculate the 1-$\sigma$ error on the amplitude using linear error propagation of the independent parameters, for which the standard deviations of $N_\gamma$ and $B$ are calculated as the square root of the considered parameter.

If none of the detection criteria are passed, an upper limit on the oscillation amplitude is determined with equation \ref{amp} using the strongest (average) signal power ($Z^2_s$) from those that did not pass the detection criterion (non-significant signals). This definition results in upper limits that are defined in the exact same manner as the detections, such that the results from the detections can be compared to those of the non-significant signals. Note that \citet{muno2004} defined the upper limit as the largest amplitude that could be obtained from the first five seconds after the start of the burst decay, because they argue that most detections are found in this phase. However, this upper limit is not necessarily based on the largest measured power found throughout the whole burst, and their non-detections are thus defined differently from the detections.

\begin{figure*}
	\begin{center}
	\includegraphics[width=0.75\textwidth]{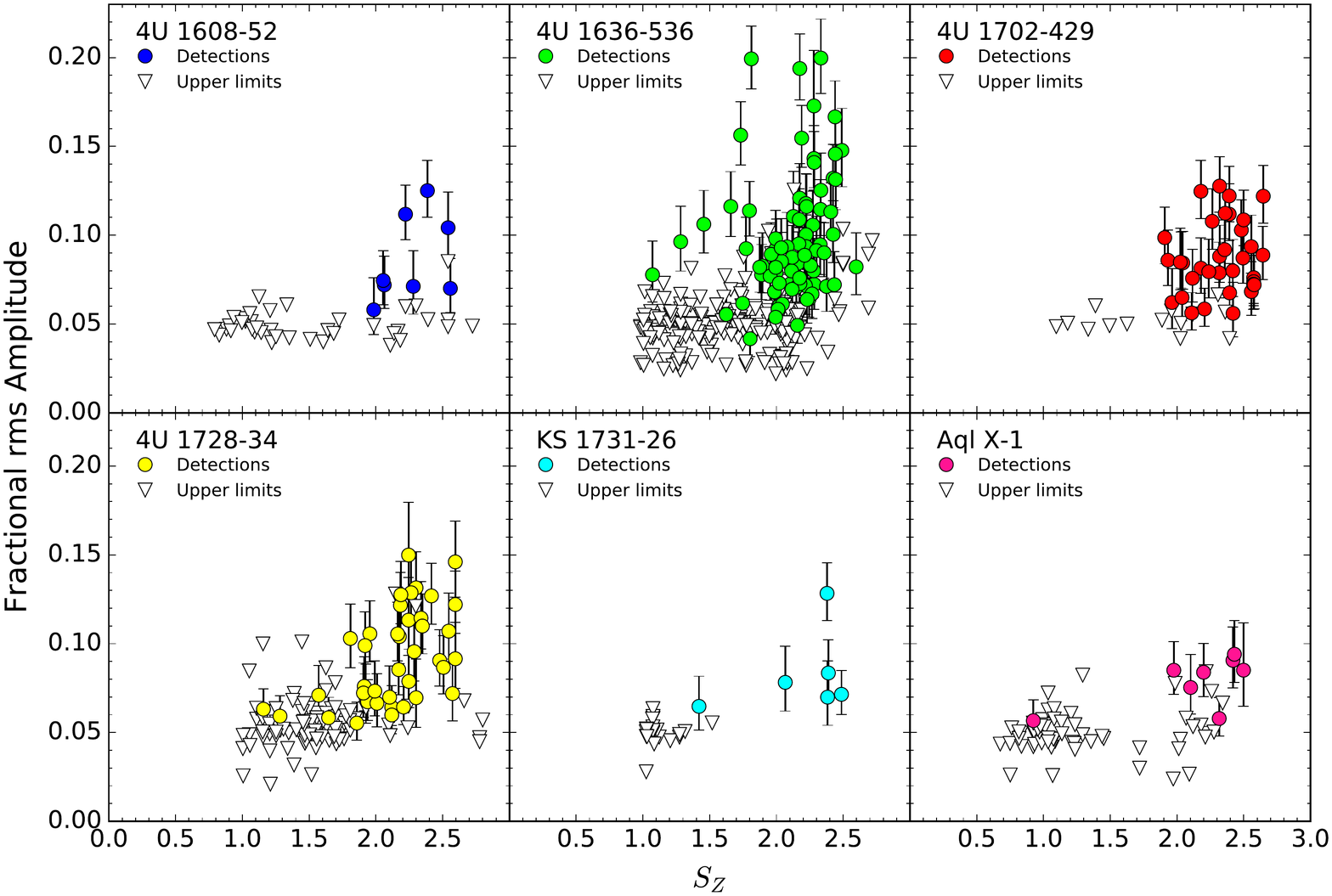}
	\end{center}
	\caption{The accretion rate (indicated by $S_\text{Z}$) dependence of burst oscillation amplitude for each of the six oscillation sources. From each analysed burst we selected the signal with the largest signal power and plotted, depending on whether or not the signal passed the detection criterion, the corresponding amplitude (coloured circles) or amplitude upper limit (triangles) as function of $S_\text{Z}$. This figure includes only the bursts with known $S_\text{Z}$. }\label{apanel}
\end{figure*}

From the oscillation signals detected in a burst, we select the amplitude  of signal with the largest signal power to compare with the results from other bursts (see Figure \ref{141}). We thus select one specific time-frequency bin for each individual burst. If no oscillation signals are found throughout an entire burst, we select the upper limit found for the signal with the largest non-significant signal power.

\subsection{Analysis of sources without detected oscillations}\label{newsources}
The two sources without previous detected oscillations, 4U 1705-44 and 4U 1746-37, should in principle be analysed using the exact same method as the sources with oscillations, to be able to compare the results with the oscillation sources. However, the main problem is that for the two new sources, the oscillation frequency is unknown, such that we do not know in which 10 Hz frequency band to look for oscillations.  What we do is to perform the analysis described above for every 10 Hz frequency band between $50 \text{ Hz}\leq\nu\leq2050\text{ Hz}$ (the frequency windows on which we perform the analysis are thus defined as $50-60, 60-70, 70-80, ..., 2040-2050$). The frequency upper limit is a frequency that encompasses the breakup speed of all current reasonable neutron star equation of state models \citep[e.g.][]{haensel2009, watts2015}. Since the proposed oscillation mechanisms assume that the oscillation frequency is close to the spin frequency, the frequency range we trace has to be consistent with the allowed spin frequencies. As mentioned, we set a lower limit of $50$ Hz, because below this limit the expected red noise signal is very large, which would make the analysis more complex. However, it should be noted that there is one confirmed burst oscillation source, IGR J17480-2446, with a frequency of $11$ Hz \citep{cavecchi2011}. 

For each frequency window we analyse the data of the non-oscillating sources in the same way as the oscillation sources. However, since there are 199 frequency bands for which this method is applied, each burst of the non-oscillating sources is analysed 199 times. This significantly increases the chance of detecting a noise signal. The signals that pass the detection criterion set for bursts that are analysed only once, can therefore generally not be considered significant. Only when the signal is so large that it is still found to be significant when taking into account the total number of trials, can we conclude that the oscillation signal is likely not caused by noise (see Section \ref{results:newsources}). 

The advantage of this method is that we can compare the results of any frequency window to those of the oscillation sources to determine whether or not the behaviour of the non-oscillating source is anomalous. We select from both non-oscillating sources the results from the frequency band in which most signals are found that pass the detection criterion of the oscillation sources, since this is the best test of whether or not the source is anomalous.\\

\section{Results}
\label{sec:results}

\subsection{Oscillation sources}
\subsubsection{Accretion rate dependence of burst oscillation amplitude}\label{resultsacc}
In Figure \ref{apanel} the results of our analysis are plotted for each of the six oscillation sources. We plot from each burst the amplitude of the strongest oscillation signal as function of $S_\text{Z}$, which indicates the accretion state.  Most of the detected oscillations were found to pass detection criterion 1 (149 out of 185 bursts with oscillations). We do not show any errors on $S_\text{Z}$, because these are not yet available from the MINBAR database. 

\begin{figure*}
	\begin{center}
	\includegraphics[width=0.8\textwidth]{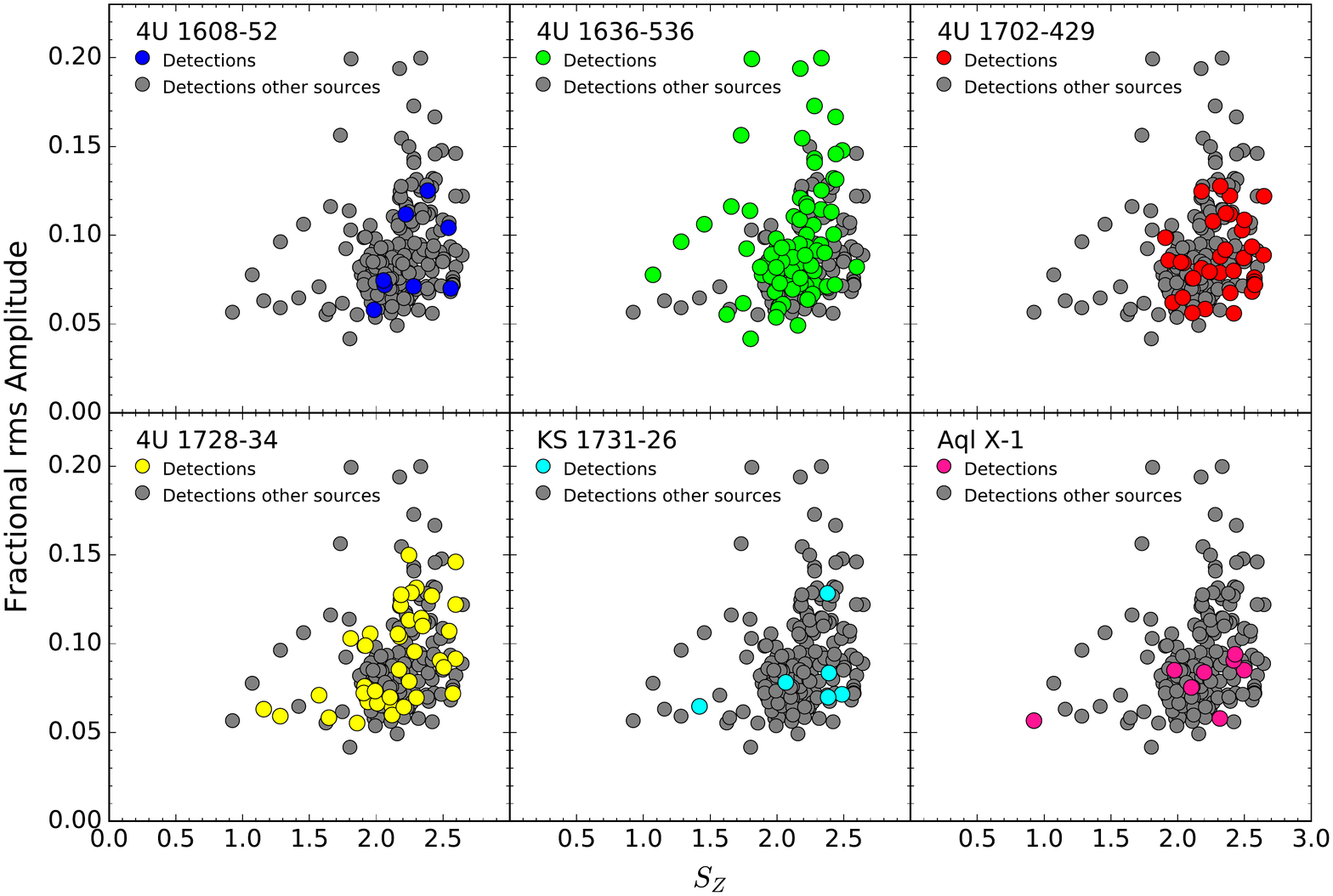}
	\end{center}
	\caption{The behaviour of the amplitudes of detected oscillations as function of $S_\text{Z}$ for each of the individual sources (coloured dots) compared to the results of detected oscillations from the other five sources combined (grey dots). Each source shows a similar trend in fractional rms amplitude as function of $S_\text{Z}$. Bursts with unknown $S_\text{Z}$ are omitted form this plot.}\label{aoverplot}
\end{figure*}

\begin{figure*}
	\begin{center}
	\includegraphics[width=0.8\textwidth]{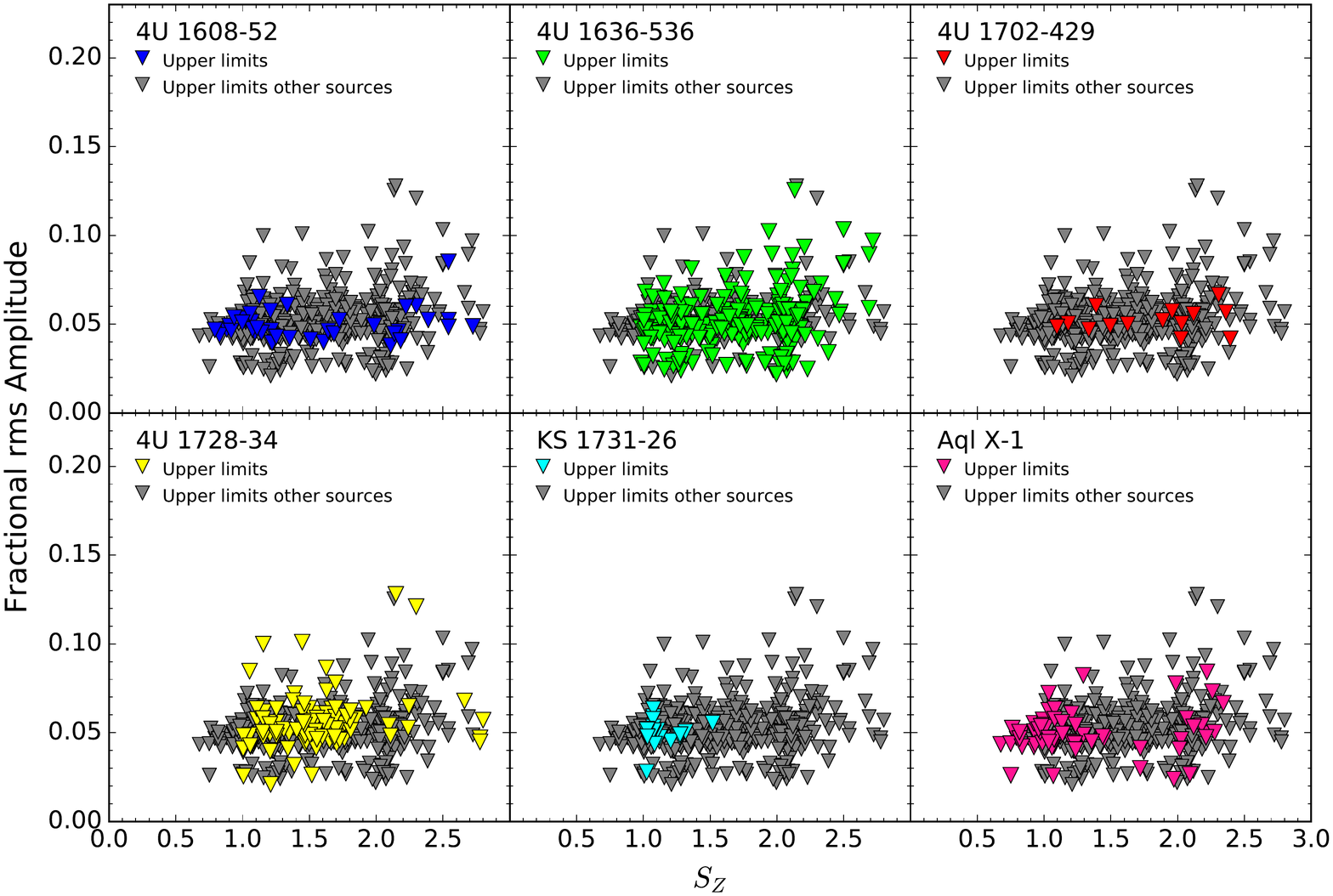}
	\end{center}
	\caption{The behaviour of the amplitude upper limits of non-detections as function of $S_\text{Z}$ for each of the individual sources (coloured triangles) compared to the amplitude upper limits from the non-detections of the five other sources combined (grey triangles). Each source shows a similar trend in fractional rms amplitude upper limit as function of $S_\text{Z}$. Bursts with unknown $S_\text{Z}$ are omitted form this plot.}
	\label{aoverplotlimits}
\end{figure*}

Our results are consistent with Figure 2 from \citet{muno2004}. There are no deviations in the general distribution of the amplitudes of the detections as function of $S_\text{Z}$. Extension of the data sample seems to cause a larger scatter of the amplitudes rather than a confinement towards any fittable trend. A more specific comparison with \citet{muno2004} shows that there does appear to be a small offset in the strongest signal of individual bursts; our amplitudes and upper limits tend to be slightly higher by $0.02-0.03$. However, we do not detect a constant offset. We will discuss this difference in more detail in section \ref{methodeffects}.  \\
\indent Figure \ref{apanel} shows that most of the detected oscillations are found at high accretion rate; $S_\text{Z}>1.7$ (see Table \ref{detectionstable} as well). However, at lower accretion rates, oscillations can be detected as well. There does not seem to be a  limit on $S_\text{Z}$ below which oscillations can no longer be detected, since for 4U 1636-536 and 4U 1728-34 detections are found in the bursts observed at the lowest $S_\text{Z}$. Furthermore, all sources show similar behaviour in oscillation amplitude as function of $S_\text{Z}$; this holds both for detected oscillations (see Figure \ref{aoverplot}) as well as for the amplitude upper limits of the non-detections (Figure \ref{aoverplotlimits}). In Figure \ref{aoverplot} (Figure \ref{aoverplotlimits}) we plot the results of the detections (non-detections) of the individual sources on a background formed by the results of the detections (non-detections) from the other five investigated oscillation sources. The distribution of results from individual sources are consistent with the distribution of the general population of the other sources.  

There is a trend in fractional rms amplitude as function of $S_\text{Z}$. The detections found at low accretion rate ($S_\text{Z}\leq1.7$) generally have low fractional rms amplitudes, $A_\text{rms}\leq0.10$. The amplitude upper limits of the non-detections are equally low. At higher accretion rate, the signals are found to have amplitudes over a broad spread of amplitudes; $0.05\leq A_\text{rms}\leq0.20$. A significant fraction of oscillation signals has a large amplitude: $\sim1/3$ of the oscillations have an amplitude $A_\text{rms}> 0.10$ at $S_\text{Z}>1.7$. The only exception is Aql X-1, in this source all six bursts with detected oscillation signals at $S_\text{Z}>1.7$ are found to have amplitudes that satisfy $A_\text{rms}\leq0.10$. However, it should be noted that the number statistic is rather low.

\begin{deluxetable}{lll}
\tablecaption{Detectability for high and low $S_\text{Z}$}
\tablehead{\colhead{Source} & \multicolumn{2}{l}{Bursts with detected oscillations} \\ 
\colhead{} &  \colhead{$S_\text{Z}\leq1.7$} & \colhead{$S_\text{Z}>1.7$}} 
\startdata
4U 1608-52  &  0/27  &  8/20  \\
4U 1636-536  &  7/141  &  73/196  \\
4U 1702-429  &  0/6  &  35/43  \\
4U 1728-34  &  4/76  &  36/55  \\
KS 1731-26  &  1/22  &  5/5  \\
Aql X-1  &  1/50  &  7/21
\enddata
\tablecomments{Fraction of the bursts in which oscillations have been detected for high ($S_\text{Z}>1.7$) and low ($S_\text{Z}\leq1.7$) accretion rates. Note that we omit bursts for which the $S_\text{Z}$ value is unknown.}
\label{detectionstable}
\end{deluxetable}

\begin{figure*}
	\begin{center}
	\includegraphics[width=0.74\textwidth]{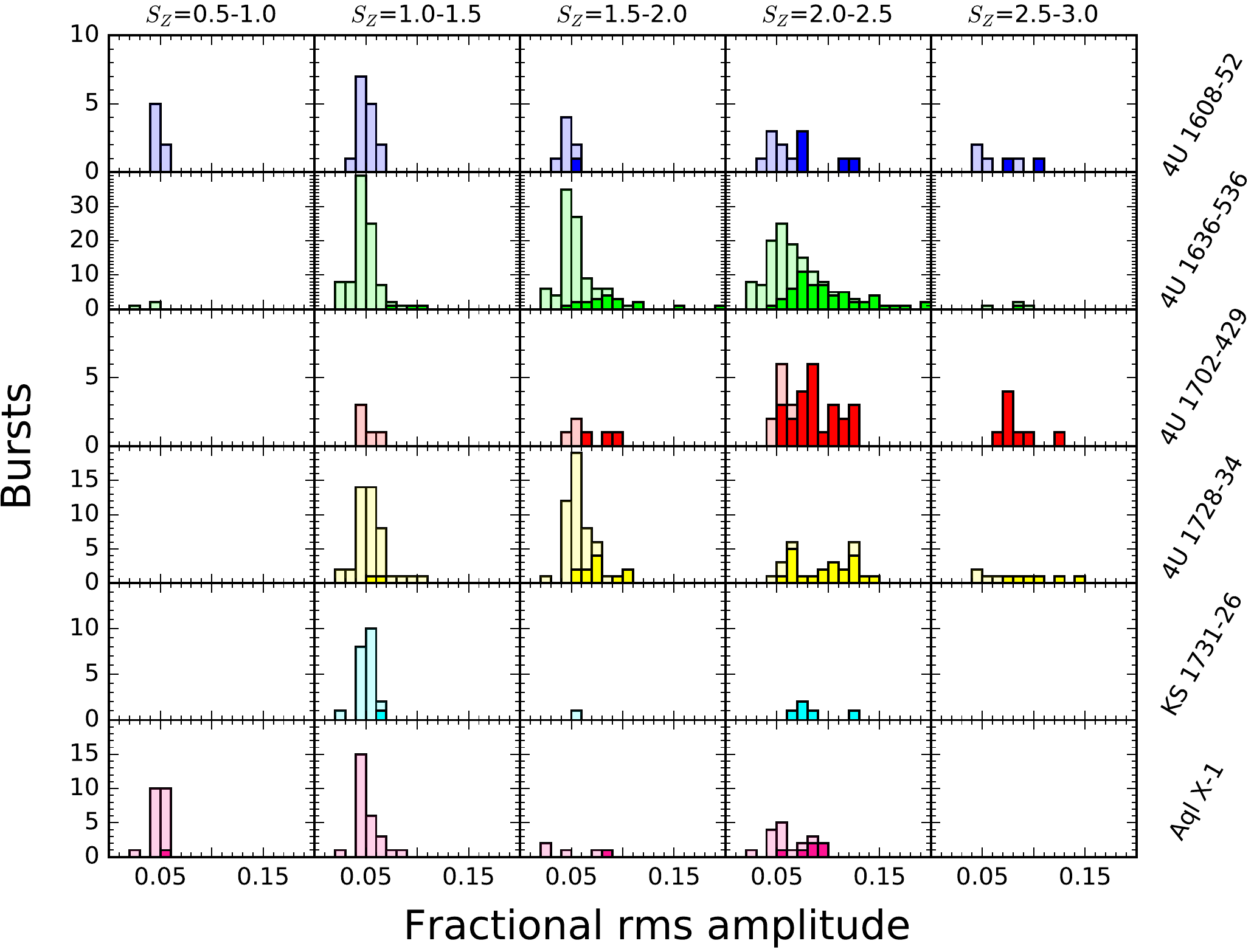}
	\end{center}
	\caption{Amplitude distribution for different ranges of accretion rate. The full range of accretion rates over which burst have been observed is divided into five equally spaced $S_\text{Z}$ bins (vertical panels). For each source (rows), a histogram of the spread in amplitudes of both the detections and non-detections is shown for each of the accretion rate ranges. The light areas in each histogram show the results of the combination of detections and non-detections, while the dark areas indicate the spread in detected oscillation amplitudes only. The peak of the full distribution seems to be around $A_\text{rms}=0.05$ at all accretion rates and in all sources, but as the accretion rate increases, the distribution significantly broadens. We note that each source has a different vertical axis and that this analysis only includes the bursts with known $S_\text{Z}$.}\label{histograms}
\end{figure*}

The distribution in amplitude as function of $S_\text{Z}$ is emphasised in Figure \ref{histograms}, which shows for each source (horizontal panels) histograms of the distribution of amplitudes for different $S_\text{Z}$ ranges (vertical panels). This figure includes the distribution of amplitudes of oscillations in combination with amplitude limits of non-detections. 4U 1636-536 and 4U 1728-34 have the largest burst sample and therefore the most reliable results. In the range $1.0\leq S_\text{Z}\leq1.5$ the amplitude distribution is strongly peaked around $A_\text{rms}=0.05$. On the other hand, at higher accretion rate ($2.0\leq S_\text{Z}\leq2.5$) the peak of the distribution is still at 0.05, but the amplitude distribution is significantly broader due to the contribution of amplitudes of detected oscillations.  Note that we tried but were unable to find a simple functional fit to the distributions.

\subsubsection{The dependence of oscillation detectability on burst phase}
The proposed burst oscillation models differ in expected oscillation amplitudes and burst phases (rise, peak, or tail) during which oscillations can be detected. To be able to explain the results in the light of oscillation models, we also analysed during which burst phase the detections are found using the obtained analysis figures of the individual bursts. We focus on the burst phase during which detections are found in bursts at both low accretion rates ($S_\text{Z}\leq1.7$) and in bursts at high accretion rates. At high accretion rate we specifically focus on oscillations signals with rms amplitudes that exceed $A_\text{rms}=0.10$, to find indications of what mechanism might cause such high amplitudes. The $S_\text{Z}$ limit is based on the results of sources 4U 1636-536 and 4U 1728-34, as there seem to be more oscillations and oscillations with higher amplitudes above this limit. Also, this limit is consistent with earlier studies by \citet{galloway2008} who stressed that more detections were found for $S_\text{Z}\geq1.75$. 

We discriminate between oscillations detected during the burst rise, peak, and tail. First we determine the maximum count rate of the burst using 0.25 second time resolution. The boundaries of the peak phase are then defined as the first and last time bin that exceed 90\% of the maximum count rate. The burst rise phase starts at the burst start time and ends at the beginning of the peak phase, and similarly, the tail starts at the end of the peak phase and ends at the last analysed time bin. This method is similar to the one described in \citet{galloway2008}. In cases where the selected 5000-count bin falls on both sides of one of the boundaries, we set the burst phase of the oscillation signal equal to the phase in which the signal was measured over the largest timespan.

4U 1636-536 has a significantly larger burst sample than all other sources, and is therefore the only source with a significant number of detections at different accretion rates. Determining the dependence of oscillation detectability on burst phase will therefore provide results with the highest significance for this source, but we present the results from all sources for completeness (see Table \ref{phase}). We note that in Table \ref{phase}, we only present the burst phase results of the selected time bin (the one with the largest signal power) in each burst and do not consider oscillation signals with smaller signal powers detected in other burst phases. 

\begin{deluxetable}{llllllllll}
\tablewidth{\columnwidth}
\tablecaption{Burst phase analysis}
\tablehead{\multicolumn{1}{l|}{Source} & \multicolumn{3}{c|}{$S_\text{Z}\leq1.7$} &\multicolumn{6}{c}{$S_\text{Z}>1.7$} \\ 
\multicolumn{1}{l|}{} & \colhead{}  & \colhead{} & \multicolumn{1}{l|}{} & \multicolumn{3}{l|}{$A_\text{rms}\leq0.10$} & \multicolumn{3}{l}{$A_\text{rms}>0.10$} \\
\multicolumn{1}{l|}{} & \colhead{R} & \colhead{P} & \multicolumn{1}{l|}{T} & \colhead{R} & \colhead{P} & \multicolumn{1}{l|}{T} & \colhead{R} & \colhead{P} & \colhead{T} }
\startdata
4U 1608-52 & 0 & 0 & 0 & 2 & 1 & 3 & 1 & 0 & 1 \\ 
4U 1636-536 & 4 & 3 & 0 & 10 & 3 & 38 & 17 & 1 & 4 \\
4U 1702-429 & 0 & 0 & 0 & 5 & 5 & 17 & 0 & 0 & 8 \\
4U 1728-34 & 0 & 1 & 3 & 4 & 7 & 10 & 7 & 3 & 5 \\
KS 1731-26 & 0 & 1 & 0 & 3 & 0 & 1 & 0 & 0 & 1 \\
Aql X-1 & 0 & 1 & 0 & 2 & 1 & 4 & 0 & 0 & 0
\enddata
\tablecomments{
This table shows how many signals were detected during either the rising phase (R), peak (P), and tail (T) of the burst. For each burst with oscillations, we only determined the phase of the strongest signal. We discriminate bursts observed at high ($S_\text{Z}>1.7$) and low accretion rate ($S_\text{Z}\leq1.7$). In burst observed at high accretion rate, we also distinguish oscillation signals with large amplitude ($A_\text{rms}>0.10$) from low amplitude oscillations ($A_\text{rms}\leq0.10$).}
\label{phase}
\end{deluxetable}

Oscillations have been detected in 81 of the 339 observed bursts from 4U 1636-536 in our sample. Seven of these bursts occurred while the source was in a low accretion state and  all of the detections in these bursts were found either during the rising phase or peak of the burst. Moreover, considering all detected oscillations throughout each of these bursts, we stress that none of these seven bursts were found to have oscillation signals in the tail of the burst. From the 22 bursts observed at $S_\text{Z}>1.7$ that were found to have oscillation signals with $A_\text{rms}>0.10$, 18 were observed during the rising phase or peak of the burst. 

\indent Overall, for 4U 1636-536 two trends seem to be present: at low accretion rate ($S_\text{Z}\leq1.7$), oscillation signals are only found during the rising phase of the bursts or during the peak rather than the tail, and at high accretion rate most of the high amplitude oscillations ($A_\text{rms}>0.10$) are detected in the rising phase. This second trend seems to be present in the bursts of 4U 1728-34 as well (see Table \ref{phase}). In general most oscillations are detected at high accretion rate and have low amplitudes. These signals are most often found during the tail of the burst (in all sources). This trend is consistent with the results found by \citet{galloway2008}.

\subsubsection{Oscillation detectability as function of accretion rate}\label{sec:detectability}
\begin{figure}[h!]
	\begin{center}
	\includegraphics[width=1\columnwidth]{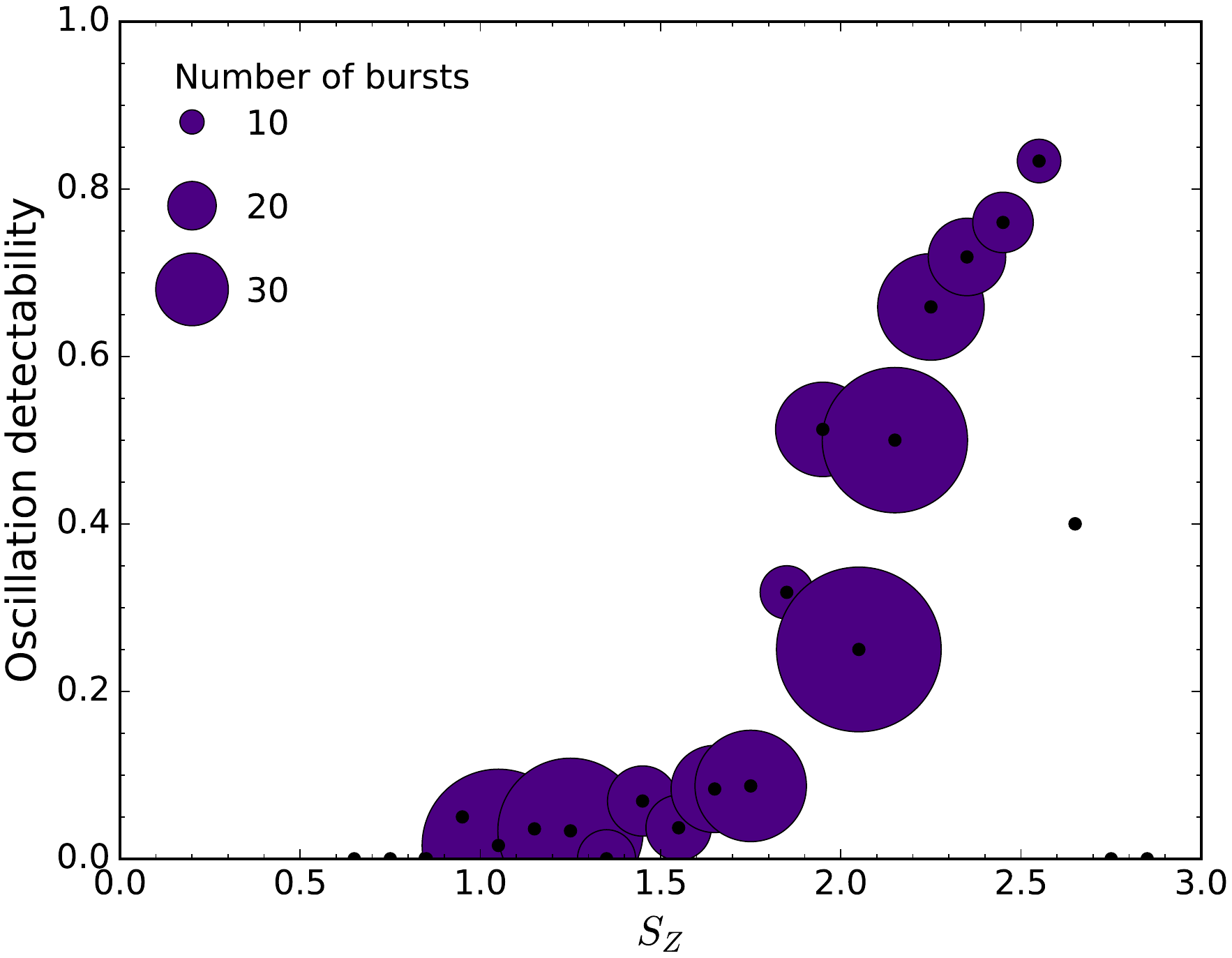}
	\end{center}
	\caption{Oscillation detectability as function of accretion rate determined from all bursts of all sources combined. The detectability is defined as the fraction of bursts with detections within non-overlapping $S_\text{Z}$ bins with 0.1 width. The size of the purple circles indicates the number of bursts in the $S_\text{Z}$ bin, with larger circles containing more bursts. Note that the size of the circle does not indicate significance.  Computation of formal error bars on the number of bursts with oscillations is not straightforward since we have three different detection criteria (see Section \ref{detectcrit}) based on distributions of powers, so we have not attempted to compute these. The figure shows that while the detectability is low at the lowest accretion rates, there seems to be a steep increase in oscillation detectability for $S_\text{Z}$ values larger than 1.7. Note that we omit bursts for which the $S_\text{Z}$ value is unknown.}
	\label{detectability}
\end{figure}

\noindent Since there are significantly more detections of oscillations found at higher accretion rate, we decided to look at the detectability of the oscillations as function of accretion rate. We combined the results from all sources, divided the $S_\text{Z}$ range $[0.6-2.8]$ into non-overlapping bins with 0.1 width and determined in each bin what fraction of the burst contained detected oscillation signals. The oscillation detectability is defined as the number of bursts with oscillations in one $S_\text{Z}$ bin, divided by the total number of bursts in the considered bin. We used the combined results rather than those of individual sources to increase the number of bursts per bin and with that the significance of each point. Combining the results is justified by the observation that the behaviour of each of the oscillation sources was found to be similar as function of accretion rate (Section \ref{resultsacc}). \\
\indent Figure \ref{detectability} shows the oscillation detectability as function of $S_\text{Z}$. For $S_\text{Z}<1.7$ the oscillation detectability is smaller than 10\%. Moreover, for $S_\text{Z}<1$ no oscillations were detected. However, because of the low number of bursts in this range (note that in Figure \ref{detectability} the number of bursts in a specific $S_\text{Z}$ range is indicated by the size of the purple circles), these points can not be considered statistically significant. For $S_\text{Z}$ values larger than 1.7, the oscillation detectability significantly increases. A steep increase of oscillation detectability as function of accretion rate can be observed for $1.7\leq S_\text{Z}< 2.5$. The oscillation detectability goes up to 85\% for $2.5\leq S_\text{Z}< 2.6$. For the highest accretion rates ($S_\text{Z}>2.6$) the detectability is found to decrease, but so does the significance.

\subsection{Sources without previously detected oscillations}\label{results:newsources}
For the two sources without previously detected oscillations, 4U 1705-44 and 4U 1746-37, we looked for signals within each 10 Hz frequency band between 50 and 2050 Hz. We thus carried out the analysis of all bursts for 199 different frequency bands. For both sources we selected the results of the frequency band with the most positive outcome based on number of signals that pass one of the criteria for the oscillation sources and the strength of the corresponding signal power.  We do this because if the properties of the signals in this band (the best candidate for a detection) are anomalous compared to the sources that do exhibit burst oscillations, then the other bands will be even more so.  

In this section we refer to signals that pass one of criteria for a single searched frequency window (these are the criteria for the oscillation sources) as pass signals. \textit{These pass signals are not detections}, because for the two sources without previous detected oscillations we apply our analysis method to 199 frequency windows. This means that if we take into account the number of trials for these two sources, none of the signals can be considered significant. The oscillation signal of a single bin detection (detection criterion 1) would have to be as strong as $Z_\text{s}^2=38.5$ for 4U 1705-44 and $Z_\text{s}^2=31.5$ for 4U 1746-37  in order to be considered significant.

\subsubsection{4U 1705-44}
\begin{figure*}
	\begin{center}
	\includegraphics[width=0.58\textwidth]{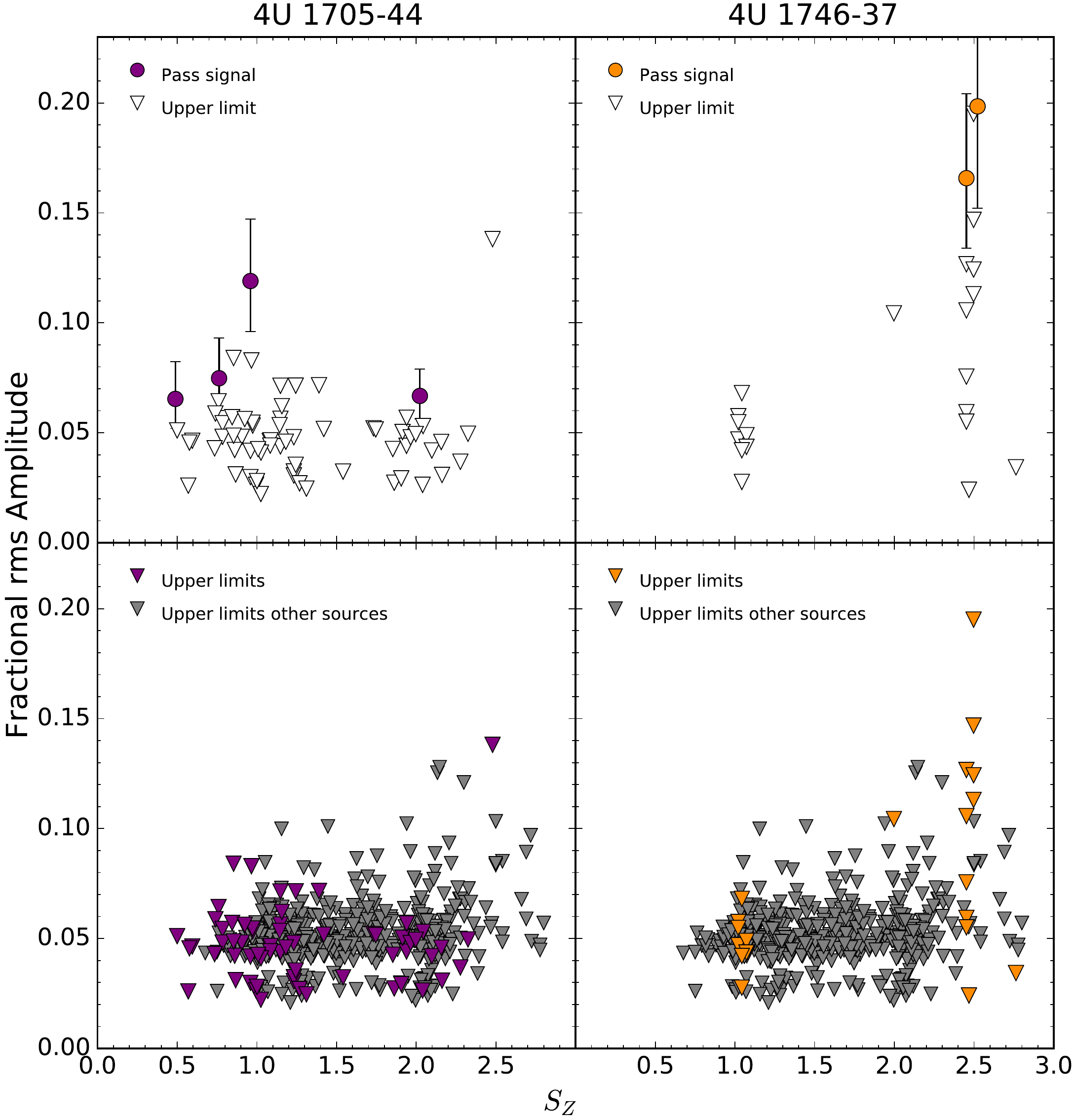}
	\end{center}
	\caption{Upper panels: Fractional rms amplitude as function of accretion rate ($S_\text{Z}$) for the two sources without previously detected oscillations. Since the spin frequency both of these sources is unknown we plot the results for the frequency range in which the most significant results were obtained. None of these results are detections. Lower panels: results of the non-oscillating sources (purple: 4U 1705-44, and orange: 4U 1746-37) plotted on a grey background formed by the results of the oscillation sources. In all panels, dots indicate pass signals and triangles represent upper limits.}\label{fignewsources}
\end{figure*}

\noindent From 4U 1705-44 a total of 70  bursts were analysed after the burst selection process (Section \ref{sample}). In the frequency range $\nu=1800-1810$ Hz we measured four pass signals. This range was selected as most positive and therefore we present the results obtained from this frequency band (see Table \ref{1705table} for the three frequency bands with the best results). The largest signal power amongst the four pass signals is $Z^2_\text{s}=22.7$. \\
\indent Figure \ref{fignewsources} shows the results of the fractional rms amplitude as function of $S_\text{Z}$ (upper left panel), as well as an overplot of the results on a background of the results from all oscillation sources (lower left panel). From the upper panel one can observe that very few bursts are observed at the highest accretion rates. The lower panel shows that the distribution in fractional rms amplitude as function of accretion rate of the observed signals seems to be similar to those of the oscillation sources. 

\begin{deluxetable}{lllll}
\tablecaption{Frequency analysis for 4U 1705-44}
\tablehead{\colhead{Rank} & \colhead{$\nu$ Band} & \colhead{Pass} & \colhead{Max $Z^2_\text{s}$} & \colhead{Significant} \\ 
\colhead{} & \colhead{(Hz)} & \colhead{} & \colhead{} & \colhead{}} 
\startdata
1    & $1800-1810$           & 4          & 22.7     & No                     \\
2    & $270-280 $            & 3          & 20.8        & No                   \\
3    & $1820-1830$           & 3          & 13.7        & No               
\enddata
\tablecomments{None of the signals meet the detection criteria and we thus do not detect oscillations for any frequency}
\label{1705table}
\end{deluxetable}

\subsubsection{4U 1746-37}
We analysed 22 bursts of 4U 1746-37 that passed the burst selection, and found in all frequency bands for which the analysis was carried out no more than one pass signal. We present the results from the frequency range $\nu=1880-1890$ Hz, because the pass signal in this range has the largest signal power; $Z^2_\text{s}=24.4$ (See Table \ref{1746table}). The upper right panel of figure \ref{fignewsources} shows the fractional rms amplitudes as function of $S_\text{Z}$. A notable observation from this figure is that more than half of the observed bursts at high accretion rate are found to have upper limits larger than $A_\text{rms}=0.10$, while in all oscillation sources the upper limits are rarely found to be this large. For most bursts of the oscillation sources, the upper limits are lower than the typical amplitudes of detected oscillations at that accretion rate, because in general a weak signal has a very low amplitude. Other than that, the selected signals from the observed bursts of this source show no abnormalities in distribution of amplitudes as function of accretion rate (lower right panel of figure \ref{fignewsources}). 

\begin{deluxetable}{lllll}
\tablecaption{Frequency analysis for 4U 1746-37}
\tablehead{\colhead{Rank} & \colhead{$\nu$ Band} & \colhead{Pass} & \colhead{Max $Z^2_\text{s}$} & \colhead{Significant}\\ 
\colhead{} & \colhead{(Hz)} & \colhead{} & \colhead{} & \colhead{}} 
\startdata
1    & $1170-1180$           & 2          & 24.4      & No    \\
2    & $1970-1980$           & 1          & 24.3      & No    \\
3    & $1240-1250$           & 1          & 22.8      & No  
\enddata
\tablecomments{None of the signals meet the detection criteria and we thus do not detect oscillations for any frequency}
\label{1746table}
\end{deluxetable}

\subsubsection{Statistical significance}\label{statsig}
To determine whether or not the two new sources are anomalous compared to the oscillation sources, we tried to obtain the probability of finding a distribution of pass signals as low as we did. Initially we tried to do this by fitting the distribution of amplitudes as function of $S_\text{Z}$ observed for the oscillation sources. However, we were unable to find a simple function that would adequately fit the data (primarily due to the broad spread in amplitudes at high accretion rate). Then we fitted the distributions of amplitudes in different $S_\text{Z}$  bins (Figure \ref{histograms}) with several functions (see Section \ref{resultsacc}).

Subsequently we applied a more robust method to obtain a value for the probability of the amplitudes found for the two sources without previously detected oscillations. Since 4U 1636-536 has the most bursts and the largest amount of bursts with oscillation signals, we set this source as model distribution. Most detections are found at high accretion rate ($S_\text{Z}>1.7$) and the amplitudes of the oscillations are higher. Larger fractional rms amplitudes indicate stronger signals, and we thus expect to find the most and the most significant results at high accretion rate. We determine how likely it is that the signals that passed any of the detection criteria, observed in bursts at high accretion rate of the two non-oscillating sources, have amplitudes as low as they do. We use the distribution formed by the detected oscillations in the bursts from 4U 1636-536 to determine this probability. First, we determine how many pass signals ($N_\text{obs}$) are found for each of the two non-oscillating sources at $S_\text{Z}>1.7$ and what the largest amplitude ($A_\text{max}$) amongst these pass signals is. Then we randomly pick $N_\text{obs}$ amplitudes of the detected signals with $S_\text{Z}>1.7$  of the model distribution, and determine whether or not the condition that all $N_\text{obs}$ amplitudes satisfy $A_\text{rms}\leq A_\text{max}$ is met. We repeat the sampling process 10000 times and determine how often the condition is satisfied. We find that for 4U 1705-44 there is a 15\% chance of finding oscillation amplitudes as low as those observed. For 4U 1746-37 this chance is 95\%.  We conclude that neither source is anomalous given the limitations of current data.  

\section{Discussion}
\label{sec:discussion}

\subsection{The accretion rate dependence of burst oscillation amplitude}
From the six oscillation sources three general trends were observed.   First, oscillations can be found at all accretion rates: there are no $S_\text{Z}$ ranges for any source in which oscillations are absent to a degree that can be considered statistically significant.  Secondly, most oscillations are found at high accretion rates. For $S_\text{Z}\leq 1.7$ oscillations are detected in less than 10\% of the bursts, while at higher accretion rates the fraction of bursts with detected oscillations rises to 85\% at $S_\text{Z}=2.5$. Thirdly, the oscillations detected in bursts at low accretion rate have low amplitudes, $A_\text{rms}\leq 0.10$, while the oscillations in bursts at higher accretion rate have a broad distribution of amplitudes ($0.05\leq A_\text{rms}\leq0.20$).  Detection limits vary somewhat from burst to burst, but in general amplitudes $A_\text{rms} \lesssim 0.05$ would not be detectable.   All of these observations are consistent with the results from \citet{muno2004}. Extension of the data sample emphasised the similarities in amplitudes as function of $S_\text{Z}$ between the different sources and allowed us to determine absolute values for oscillation detectability at different accretion rates.

We set out to determine how the oscillation amplitude changes as function of $S_\text{Z}$. We were not able to fit a smooth function through the detected amplitudes; the scatter is too large. However the distribution is also inconsistent with a step function: there is no $S_\text{Z}$ limit below which the oscillation mechanism simply switches off. 

We also investigated the bursts of two non-oscillation sources with bursts at high $S_\text{Z}$. We found that the non-detection of oscillations in these sources does not contradict the trends observed in the oscillation sources. 4U 1705-44 has very few observations at high $S_\text{Z}$, reducing the chance of detecting oscillations.  4U 1746-37 has a larger sample of bursts at high $S_\text{Z}$, but the distance to this source is much larger. While the other seven investigated sources have estimated distances $<$ 8.2 kpc, the distance to 4U 1746-37 is thought to be in the range $13-25$ kpc \citep{galloway2008}.  The fact that the bursts are fainter as a result reduces the chance of detecting oscillations. We conclude that neither source shows anomalous behaviour.

\subsubsection{Burst oscillation amplitudes in light of current theories}
\label{amptheory}
\begin{deluxetable*}{llll}
\tablecaption{Summary of model predictions}
\tablehead{\colhead{} & \colhead{Hotspot} & \colhead{Surface waves} & \colhead{Cooling wakes} }\\ 
\startdata
$A_\text{rms}$ max. & $0.2$  &$0.1$& $0.2^\text{a}$    \\ \hline
Burst phase$^*$ & R P$^\text{b}$ T$^\text{b} $ &  T&  T \\ \hline
$S_\text{Z}$ dependence &  - Ignition latitude might       &            - Signals might only be able &    \\
& depend on $S_\text{Z}$ &  to grow to detectable size at &\\
&  &  high $S_\text{Z}$&\\ \hline
Ignition latitude ($\phi_\text{ign}$) dependence& - If stationary, max. $A_\text{rms}$ lower &           & - Max. $A_\text{rms}$ decreases with \\
& for high $\phi_\text{ign}$& &  increasing $\phi_\text{ign}$\\
&- Coriolis force confinement && \\
& more effective at high $\phi_\text{ign}$, &&\\
& reducing low $\phi_\text{ign}$ amplitudes
\enddata
\tablecomments{Empty boxes indicate that at the present, there are no clear model predictions. \\
$^*$ R = rising phase, P = peak of the burst, T = burst tail\\
$^\text{a} A_\text{rms}$ up to 0.2 can only be obtained for asymmetric cooling wake models. \\$^\text{b}$ A hotspot can only produce oscillations during the peak and tail of the burst} if the hotspot is confined.
\label{modelpredict}
\end{deluxetable*}

Surface mode patterns cannot give amplitudes as high as hotspot models, since in the latter the bright spot can be restricted, physics permitting, to a smaller region of the star (either as it spreads from the ignition point or due to some mechanism like magnetic confinement).  Obtaining amplitudes $A_\text{rms} \sim 0.1$ with any of the suggested models is certainly feasible, but obtaining amplitudes as high as $A_\text{rms} \sim 0.2$ with surface mode models will be more challenging.  For canonical cooling wake models, amplitudes are predicted to be low, $A_\text{rms}\sim0.03$, while for asymmetric cooling wake models, amplitudes can be as large as $A_\text{rms}=0.10-0.20$, depending on the ignition latitude \citep{mahmoodifar2016}. An overview of the maximum amplitudes predicted for each model is given in Table \ref{modelpredict}, along with other relevant phenomena predicted for each model.

What then do the data require us to explain in terms of accretion rate dependence?    As the accretion rate rises, the mechanisms operating must permit higher maximum amplitudes, but also allow the possibility of the amplitude remaining low, since we see bursts with amplitudes that never get much above the detection limit even at high accretion rate. What then is the expectation in this regard for the different proposed mechanisms?  

For the spreading hotspot model, the most obvious dependence on accretion rate should come if ignition does indeed move off-equator to higher latitudes as the accretion rate increases \citep{cooper2007}.   This is due to variations in local accretion rate from rotational effects, which can cause burning to stabilise at the equator before it stabilises at higher latitudes.   This should occur for a relatively narrow band of accretion rates, but recall that it is not straightforward to map $S_\text{Z}$ to an absolute accretion rate.  The predicted effect on amplitude however is not immediately clear.  All other things being equal, a stationary hotspot at higher latitudes should give a lower amplitude than a hotspot on the equator \citep{muno2002amp}. However, the speed at which the flame spreads from the ignition point will also have an effect. Coriolis confinement is more effective off-equator \citep{spitkovsky2002,cavecchi2013,cavecchi2015} so an equatorial hotspot can be quickly wiped out as the flame spreads, reducing amplitude.  There may even be multiple phases of on and off equator ignition as the accretion rate changes, at the transition phases between different burning regimes \citep{maurer2008}.    The intrinsic dependence of flame speed on accretion rate (as e.g. the properties of the ocean change) and the dependence of stalling mechanisms such as magnetic confinement \citep{cavecchi2016} on accretion rate are not known.

So is the hotspot model consistent with our observations?  Let us start with the overall rise in oscillation amplitudes.    The very simplest model, where ignition location moves off-equator as accretion rate increases and flame spread does not change, would not explain the large rise in  maximum amplitude, since amplitude should fall.  For the spreading hotspot model to fit the data, either flame spread must slow significantly as ignition moves off-equator due to Coriolis effects (as predicted), or ignition is actually moving back on-equator at high accretion rates, due to presence of multiple burning regimes.  The alternative is that some stalling mechanism becomes more effective as accretion rate rises.   What about the spread in amplitudes seen at higher accretion rates?  In the simple spreading hotspot model, ignition latitude and flame spread speed from that point should vary monotonically with accretion rate.  It is therefore hard to see how such a broad spread is possible if this is the only mechanism in operation, unless ignition location depends on other factors as well.  

Predictions for the accretion rate dependence of surface mode amplitudes centre around the need to excite the oscillations to detectable amplitudes.  \citet{narayan2007} have argued that there is a nuclear burning driven instability mechanism that would  operate to pump mode amplitudes primarily in helium-dominated bursts, which occur preferentially at higher accretion rates.  If this is correct, surface mode amplitudes should rise as accretion rate increases.  This matches the observation of a general rise in amplitude, although surface mode models would still struggle to explain the very highest amplitudes seen at the highest accretion rates.  Given that there are mechanisms that could also suppress surface mode amplitudes again e.g. in bursts where a large convective layer develops \citep{cooper2008}, surface mode models could also viably explain the lower amplitudes seen in some bursts as well.   

\begin{figure*}
	\begin{center}
	\includegraphics[width=0.48\textwidth]{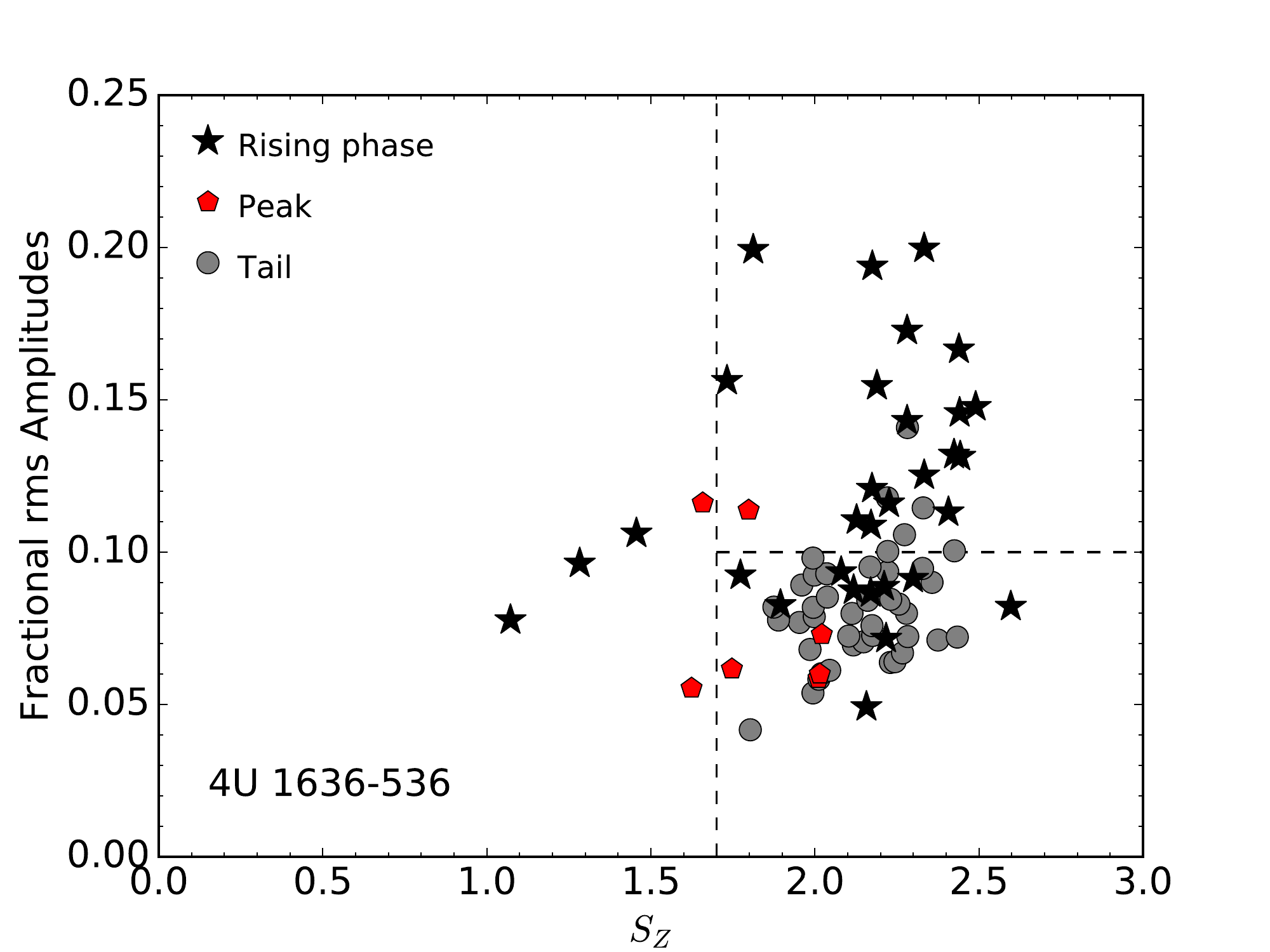}\includegraphics[width=0.48\textwidth]{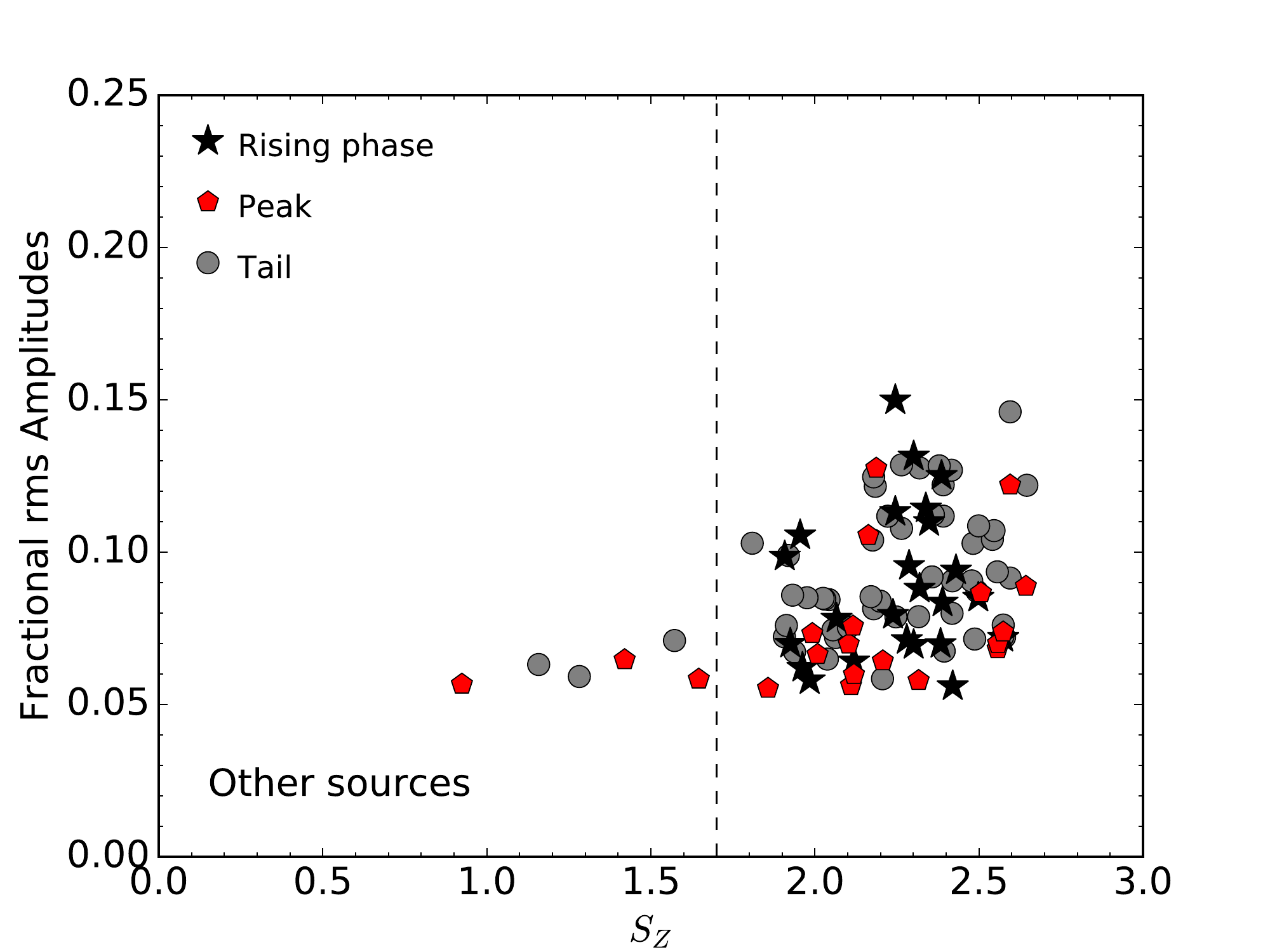}
	\end{center}
	\caption{Left: Result of the burst phase analysis for bursts with detected oscillations in 4U 1636-536. We show for each burst in which an oscillation was detected the amplitude as function of accretion rate, while the symbol indicates whether the oscillation was detected during the rising phase (black stars), peak (red hexagons), or tail (grey dots) of the burst. The vertical line indicates the separation between high and low $S_\text{Z}$, while the horizontal line indicates the separation between high and low amplitude. Note that the symbol indicates only the burst phase in which the strongest signal was found. Right: same as left panel, but for the detected oscillations in 4U 1608-52, 4U 1702-429, 4U 1728-34, KS 1731-26 and Aql X-1 combined. Since amplitudes of different sources are combined in this figure without taking into account any inclination effects, we do not distinguish high and low amplitudes.}\label{phasemodel}
\end{figure*}

Our conclusions at this stage are rather tentative. Hotspot models can in principle give rise to the highest amplitudes seen, and it is possible to explain the overall rise in the envelope of the amplitude distribution with accretion rate provided that the flame spread speed varies strongly with latitude (for simple expectations of the change in ignition latitude with accretion rate).  However, this simple model on its own would struggle to explain the fact that at similar accretion rates there are also bursts that do not show high amplitude oscillations.  Perhaps ignition location varies more than expected, or we are seeing the effects of a stalling mechanism, or flame spread speed simply varies much more than indicated by current studies.  Alternatively we are seeing a second mechanism coming into play at higher accretion rates, to give the lower amplitude population. In this scenario the high and low amplitude oscillations at high accretion rate would be caused by two different mechanisms.  Surface mode models, which can explain amplitudes up to about 0.1, would fit well in this regard since instability mechanisms that might pump them to detectable amplitudes are only expected to kick in as accretion rate rises.  We cannot comment on the cooling wake model of \citet{mahmoodifar2016} since at this stage there are no concrete predictions for accretion rate dependence against which to compare.

\subsection{The accretion rate dependence of oscillations at different burst phases}

To shed more light on the mechanisms at work, we analysed the burst phases in which oscillations are detected for bursts at different accretion rates. Figure \ref{phasemodel} shows for each burst in which oscillations were detected, the amplitude of the strongest oscillation signal as function of $S_\text{Z}$, as well as the burst phase in which this signal was detected. The left panel shows these results for 4U 1636-536 (the source with the largest burst sample), and the right for the remaining five oscillation sources combined. 

Taking the sample as a whole, the maximum oscillation amplitude can occur, for any $S_\text{Z}$, during any phase of the burst.  The very largest amplitudes seen, for a given $S_\text{Z}$, are always found in the rise.  However this latter finding depends entirely on the inclusion of 4U 1636-536 in the sample.  At the lower end of the maximum amplitude distribution envelope, all burst phases are represented (the picture for $S_\text{Z} < 1.7$ appears to show some separation, but there are too few bursts with oscillations here to draw reliable conclusions).   Interestingly however maximum amplitudes at any specific burst phase show the same general trend:  the maximum of the envelope that can be reached rises as accretion rate increases, while the lower end of the envelope remains at the detection threshold.

How do these observations fit within the context of current theoretical expectations?    A spreading hotspot is expected to cause oscillations during the rising phase of the bursts only, since the brightness asymmetry disappears as soon as the flame has engulfed the entire surface of the neutron star \citep{chakraborty2014}. A stalled hotspot would cause detectable signals in all phases of the bursts (rise, peak, and tail), as long as it is offset from the rotational pole. Since surface waves require time to grow to significant size, perhaps comparable to the duration of a burst \citep{narayan2007}, surface wave oscillations are expected in the tails only. However, it is unknown when in the tail these oscillations are expected to be strongest.   Cooling wake oscillations are only able to explain burst tail oscillations. \citet{mahmoodifar2016} calculated the expected amplitude evolution during a burst and found for canonical cooling wakes, that the amplitudes are strongest $\sim$1 second after the peak of the burst after which they would decay. For asymmetric cooling wakes the amplitude evolution of the oscillation signal is expected follow a smooth parabolic evolution, with a maximum $\sim$5 seconds after the peak of the burst. The exact amplitude evolution depends, among on factors, on the ignition latitude.   An overview of the burst phase in which oscillations are expected for each model can be found in Table \ref{modelpredict}. 

The fact that the very largest maximum amplitudes ($A_\text{rms}\geq 0.15$ are seen the rise is consistent with the idea that these are caused by a (spreading) hotspot with an appropriate observing geometry. The lower amplitude oscillations are consistent with any of the models put forward, although as stated in Section \ref{amptheory} modes would have to be excited much more quickly than current models predict if they are to explain rise phase oscillations.  However our data do now pose a strong constraint on all models.  The mechanism(s) in operation must (1) be capable of permitting higher amplitude oscillations at all burst phases as accretion rate increases and (2) still permit the generation of low amplitude oscillations at all burst phases.  Whether this can be achieved with one mechanism alone is an interesting question.   

\subsection{Method effects and assumptions}\label{methodeffects}
In this research we followed the method from \citet{muno2004} for the analysis and detection of oscillations. There are however, two main differences between our methods. First of all, \citet{muno2004} used one second time bins, while we use time bins with equal number of counts (5000 counts per bin). For a typical burst, the bins will on average have a width of one second. However, for fainter bursts, the time bins grow in width. The disadvantage of this method is that for broad bins, the signal might drift in frequency within the bin, making the signal hard to detect, even though we also look for signals in overlapping frequency bins. Also, the number of bins significantly falls for bursts with low count rate, decreasing the chance of a detection as the number of trials for the burst is lower. We excluded bursts for which the average count rate was too low to have multiple bins of reasonable width, to reduce the number of bursts for which it is very likely that no oscillations are found due to the methodology. The benefit of this method is that, since the amplitude error depends on the number of counts in a time bin, all errors will have the same contribution of the uncertainty in counts per time bin. 

As mentioned in Section \ref{resultsacc}, our oscillation amplitudes tended to be slightly higher ($\sim$$0.02-0.03$) compared to the results of \citet{muno2004}. However, an offset was only present in some bursts and the offset is not constant. We checked the results when using one second time bins, and found no deviations in the amplitudes from \citet{muno2004} in that case, indicating that the observed differences in amplitudes are caused the difference in the definition of time bins. Comparison of individual bursts showed that many of the 5000-count bins in our research are smaller than one second, especially near the peak of the burst. As a result, the selected signal of a burst often originates from a bin smaller than one second. Within smaller time bins, the strength of the oscillation signal will be less affected by frequency drifts. Consequently, the oscillation signal will be stronger. It should be noted that for bursts where the width of the time bins deviates from the assumed one second, the total amount of bins increases. This affects the probability of a signal being caused by noise. We checked the number of time bins per bursts after the analysis, and corrected signals that were wrongly classified due to the standard probability calculation (based on 16 time bins per burst) by hand.

A second difference from the method of \citet{muno2004} is that they select the upper limit of a burst oscillation as being the signal with the largest non-significant amplitude from the first five seconds after the start of the burst decay, while we select the signal with the largest non-significant power throughout the whole burst. This upper limit is computed by the same method used to obtain the oscillation signal from a single burst, such that amplitude upper limits and detected oscillation amplitudes can be compared with each other. 

We checked that a similar trend can be found in the measured powers as in the rms amplitude as function of $S_\text{Z}$. This indicates that the results do not depend on the method that we use to convert measured power into signal power and consequently into rms amplitude. A recent study showed that the persistent luminosity during a burst is variable, while we estimate a constant background from the pre-burst emission \citep{worpel2015}. This would influence the background correction that we apply while obtaining the rms amplitude. The effect found by \citet{worpel2015} is a factor of a few increase in persistent emission during the rise and peak of the burst. Consequently, if the background increase was taken into account, the measured burst oscillation amplitudes during the rise and peak of the burst would have been larger (Equation \ref{amp}). Burst tail oscillations would be unaffected. Future study would be required to determine whether such changes in persistent emission would be sufficient to give rise to the trends that we observe without requiring any change in burst oscillation mechanism.

In this research we chose to use non-overlapping time bins and to select from each burst the signal with the largest power. However, we also could have taken the average amplitude of the oscillation signal over the whole burst instead. Another method that is often used for analysis of burst oscillations is to determine the dynamical power spectrum of each burst, which requires the use of overlapping time bins. This method has the advantage that the frequency drift can be determined. 

Additionally, we chose to use the $S_\text{Z}$ value as measure of accretion rate. It should be pointed out that it is still under debate whether or not this is indeed the best measure of accretion rate. One might also consider using $\gamma$, the fraction of persistent flux compared to the Eddington flux. 

Any possible effects of rotational and binary inclination on the determined oscillation amplitude are not taken into account. One may assume that as a consequence of mass accretion, the rotational and binary inclination are aligned \citep{hills1983,bhattacharya1991,guillemot2014}, such that the effects of rotational and binary inclination can be discussed together. None of the sources are known dippers, which limits the binary inclination angles to (typically) $<75^\circ$ \citep[see e.g.][]{galloway2016}. Inclination is expected to affect the observed amplitudes originating from hotspot oscillations, depending on the ignition latitude. \citet{muno2002amp} showed that for a stationary hotspot, the maximum observed amplitude is observed for inclination angle between the observer and the neutron star spin axis: $i=180^\circ-\phi_\text{ign}$. Detailed analysis of possible correlations between the inclination angle and amplitudes is beyond the scope of this paper. Once strong constraints on the inclination angle of all systems are obtained, one can consider the possible effects on our results.

We estimated confidence limits on the maximum RMS amplitude by assuming that the measured amplitudes are distributed uniformly between zero and some maximum value, $A_\text{rms,max}$. This assumption must be viewed with caution, both because the observed distribution is far from uniform (e.g. Fig. \ref{histograms}), and also because arbitrarily small amplitudes cannot be measured, due to the detection threshold, so that the effective minimum amplitude is $>0$. Nevertheless, the largest impact on the probability density function of $A_\text{rms,max}$ is the measurements with the largest amplitudes. We find for individual sources that the highest-probability value of $A_\text{rms,max}$ was between 10 and 17\%, with the extreme sources being Aql X-1 and 4U 1636-536, respectively. For the entire sample, the inferred $A_\text{rms,max}=14.6_{-0.4}^{+0.5}$\%, with a 95\% upper limit of 15.4\%

Finally, in this research we only investigated the bursts from six of the 17 confirmed oscillation sources and two of the 87 type I X-ray burst sources without previously detected oscillations. One may wonder if the results are not biased because we only selected a subsection of the available sources. Preferably, we would investigate all oscillation sources, but the main problem with the other sources is that in those sources usually very few bursts with oscillations have been observed and typically only a small range of accretion rate. This impedes the construction of a colour-colour diagram from which $S_\text{Z}$ values can be obtained. Using a different measure of accretion rate would circumvent a possible bias that could be present due to the fact that we select only sources that exhibit large variation in accretion rate.

\indent For sources without previously detected oscillations, as similar argument holds. It would indeed be better to investigate all sources without previously detected oscillations, but we chose these two because a significant amount of bursts was observed from the selected two sources over a broad range of accretion rates, and specifically because they where observed at $S_\text{Z}>1.7$ where most oscillations are detected. The sources were therefore prime candidates for the detection of oscillation signals, and what we constrained is that at least these sources do not seem to be anomalous in their behaviour as far as can be observed. Additional monitoring of the bursts of the remaining non-oscillating sources is suggested for further research. 

\section{Conclusions}
\label{sec:conclusion}
We have analysed 765 type I X-ray bursts of six different LMXBs that are all known to exhibit thermonuclear burst oscillations. The goal of this research was to investigate the accretion rate dependence of burst oscillation amplitude in order to gain insight on the mechanism that causes the asymmetric brightness patterns on the burning surfaces of accretion neutron stars that underlie the burst oscillations. Additionally, we analysed the bursts of two sources without previously detected burst oscillations (124 bursts in total) that have been observed over a broad range of accretion rates. These two sources are observed at accretion rates $S_\text{Z}>1.7$, a limit above which most oscillations are detected. We have found the following results:
\begin{enumerate}
\item Oscillations can be detected at all accretion rates; there is no evidence for a cutoff in accretion rate below which oscillations are no longer detectable. 
\item Significantly more detections as a fraction of number of bursts are found at high accretion rate, than at low accretion rate. This confirms earlier results by \citet{muno2004}. These results become most evident from the oscillation detectability (the fraction of bursts observed in an accretion rate range in which we detected burst oscillations when combining the results of all sources). We found that for $S_\text{Z}<1.7$ the oscillation detectability is low ($\sim 10\%$), while for $1.7\leq S_\text{Z}<2.5$ the oscillation detectability steeply increases with accretion rate, up to $\sim 85\%$ at $2.5\leq S_\text{Z}<2.6$.
\item All sources show a similar distribution of oscillation amplitudes as function of accretion rate. At low accretion rates, the oscillation amplitudes are generally low; $A_\text{rms}\leq0.10$, while at higher amplitudes both high and low amplitudes are found. 
\item Analysis of the phase of the burst (rise, peak, or tail) in which we detect oscillation signals shows that oscillations can be detected in all phases of the burst.   In the best-sampled source, 4U 1636-536, we found that at low accretion rates, none of the bursts showed oscillations in the tail.  At high accretion rate, bursts with low amplitude oscillations were found in all phases of the bursts, while the high amplitude oscillations were primarily detected during the rising phase or peak of the bursts.
\item The two sources without previously detected oscillations that we studied do not seem to be anomalous in their behaviour. The lack of detected signals can be attributed to poor sampling of the high accretion state in the case of 4U 1705-44. For 4U 1746-37 the upper limits on the signals are higher than typical detected oscillation signals ($A_\text{rms, lim}\geq0.10$), which means that for this source only very strong signals would be detectable. The high upper limits are most likely caused by the fact that the distance to this source is significantly larger than the distances of the other seven sources in this research. 
\end{enumerate}

There are three types of models that try to explain burst oscillations: hotspot models, surface wave models, and cooling wake models. Oscillations of these models can be distinguished by the burst phase in which they are detected: spreading hotspot oscillations are expected to be observed during the rising phase, confined hotspots can be detected in any burst phase, and global modes and cooling wakes are expected to cause oscillations during the tail of the bursts only. The amplitudes of hotspot oscillations are in general expected to be larger than those of surface wave and cooling wake oscillations. For all models accretion rate dependence is expected.  Ignition latitudes and flame spread speeds will vary as accretion rate changes, for example, affecting rise phase oscillation detectability, and surface modes may only be excited to detectable amplitudes during the type of bursts expected at higher accretion rates. 

We see no evidence for sharp changes in the distribution of amplitudes with increasing $S_\text{Z}$, which might indicate a mechanism switching on or off.  Instead we see things change relatively continuously, implying that the same mechanisms are active but change gradually in terms of the amplitudes of oscillations that they produce.  We also see a spread of amplitudes, implying a dependence on other physical parameters in addition to accretion rate.  However, we can draw some tentative conclusions.

At all accretion rates the highest amplitudes are seen in the rising phase of the bursts, as might be expected from a spreading hotspot. The rise phase amplitudes increase as accretion rate rises, suggesting that any change in ignition latitude, which on its own might be expected to reduce amplitude, is being offset by increased confinement of the flame or a more effective stalling mechanism.  Many bursts however have their maximum amplitude in the peak or tail: and for these too the maximum amplitude that can be reached rises smoothly with accretion rate.  This puts new constraints on models for the mechanisms that may operate in the tail:  stalled hotspots, cooling wakes, or surface modes.   Further theoretical work is now required to connect rise and tail mechanisms, and to explore in more detail the predictions for accretion rate dependence of the various mechanisms. Both the overall smooth rise in amplitudes with accretion rate found in our study, and the remaining spread, should be addressed.  

\section*{Acknowledgments}
LO and RW acknowledge support from a NWO Top Grant, module
1, awarded to RW. AW acknowledges support from ERC Starting Grant No. 639217 CSINEUTRONSTAR.  DKG. acknowledges the support of an Australian Research Council Future Fellowship (project FT0991598).
This paper utilizes preliminary analysis results from the Multi-INstrument Burst ARchive (MINBAR), which is supported under the Australian Academy of Science's Scientific Visits to Europe program, and the Australian Research Council's Discovery Projects and Future Fellowship funding schemes.

\bibliography{Burst_oscillations.bbl}

\begin{thebibliography}{}
\expandafter\ifx\csname natexlab\endcsname\relax\def\natexlab#1{#1}\fi

\bibitem[Berkhout 
\& Levin(2008)]{berkhout2008} Berkhout, R.~G., \& Levin, Y.\ 2008, \mnras, 385, 1029 

\bibitem[Bhattacharya \& van den Heuvel(1991)]{bhattacharya1991} Bhattacharya, D., \& van den Heuvel, E.~P.~J.\ 1991, \physrep, 203, 1 

\bibitem[Bildsten(1998)]{bildsten1998} Bildsten, L.\ 1998, NATO 
Advanced Science Institutes (ASI) Series C, 515, 419 

\bibitem[Buccheri et al.(1983)]{buccheri1983} Buccheri, R., Bennett, K., Bignami, G.~F., et al.\ 1983, \aap, 128, 245 

\bibitem[Casella et al.(2008)]{casella2008} Casella, P., 
Altamirano, D., Patruno, A., Wijnands, R., 
\& van der Klis, M.\ 2008, \apjl, 674, L41 

\bibitem[Cavecchi et al.(2011)]{cavecchi2011} Cavecchi, Y., Patruno, 
A., Haskell, B., et al.\ 2011, \apjl, 740, L8 

\bibitem[Cavecchi et al.(2013)]{cavecchi2013} Cavecchi, Y., Watts, 
A.~L., Braithwaite, J., \& Levin, Y.\ 2013, \mnras, 434, 3526 

\bibitem[Cavecchi et al.(2015)]{cavecchi2015} Cavecchi, Y., Watts, 
A.~L., Levin, Y., \& Braithwaite, J.\ 2015, \mnras, 448, 445 

\bibitem[Cavecchi et al.(2016)]{cavecchi2016} Cavecchi, Y., Levin, Y., Watts, A.~L., \& Braithwaite, J.\ 2016, \mnras, 

\bibitem[Chakrabarty et al.(2003)]{chakrabarty2003} Chakrabarty, D., 
Morgan, E.~H., Muno, M.~P., et al.\ 2003, \nat, 424, 42 

\bibitem[Chakraborty 
\& Bhattacharyya(2014)]{chakraborty2014} Chakraborty, M., \& Bhattacharyya, S.\ 2014, \apj, 792, 4 

\bibitem[Cooper 
\& Narayan(2007)]{cooper2007} Cooper, R.~L., \& Narayan, R.\ 2007, \apjl, 657, L29 

\bibitem[Cooper (2008)]{cooper2008} Cooper, R.~L.\ 2008, \apj, 684, 525 

\bibitem[Cumming \& Bildsten(2000)]{cumming2000} Cumming, A.
\& Bildsten, L.\ 2000, \apj, 544, 453

\bibitem[Galloway et al.(2008)]{galloway2008} Galloway, D.~K., Muno, 
M.~P., Hartman, J.~M., Psaltis, D., 
\& Chakrabarty, D.\ 2008, \apjs, 179, 360 

\bibitem[Galloway et al.(2016)]{galloway2016} Galloway, D.~K., Ajamyan, A.~N., Upjohn, J., \& Stuart, M.\ 2016, arXiv:1607.00074 

\bibitem[Guillemot \& Tauris(2014)]{guillemot2014} Guillemot, L., \& Tauris, T.~M.\ 2014, \mnras, 439, 2033 

\bibitem[Haensel et al.(2009)]{haensel2009} Haensel, P., Zdunik, J.~L., Bejger, M., \& Lattimer, J.~M.\ 2009, \aap, 502, 605 

\bibitem[Hasinger 
\& van der Klis(1989)]{hasinger1989} Hasinger, G., \& van der Klis, M.\ 1989, \aap, 225, 79 

\bibitem[Heyl(2004)]{heyl2004} Heyl, J.~S.\ 2004, \apj, 600, 939 

\bibitem[Hills(1983)]{hills1983} Hills, J.~G.\ 1983, \apj, 267, 322 

\bibitem[Homan et al.(2010)]{homan2010} Homan, J., van der Klis, 
M., Fridriksson, J.~K., et al.\ 2010, \apj, 719, 201 

\bibitem[Jahoda et al.(1996)]{jahoda1996} Jahoda, K., Swank, J.~H., Giles, A.~B., et al.\ 1996, \procspie, 2808, 59 

\bibitem[van der Klis(1989)]{vanderklis1989} van der Klis, M.\ 1989, NATO Advanced Science Institutes (ASI) Series C, 262, 27 

\bibitem[Mahmoodifar 
\& Strohmayer(2016)]{mahmoodifar2016} Mahmoodifar, S., \& Strohmayer, T.\ 2016, \apj, 818, 93 

\bibitem[Maurer 
\& Watts(2008)]{maurer2008} Maurer, I., \& Watts, A.~L.\ 2008, \mnras, 383, 387 

\bibitem[M{\'e}ndez et al.(1999)]{mendez1999} M{\'e}ndez, M., van der Klis, M., Ford, E.~C., Wijnands, R., \& van Paradijs, J.\ 1999, \apjl, 511, L49 

\bibitem[Muno et al.(2002a)]{muno2002a} Muno, M.~P., Chakrabarty, 
D., Galloway, D.~K., \& Psaltis, D.\ 2002a, \apj, 580, 1048 

\bibitem[Muno et al.(2002b)]{muno2002amp} Muno, M.~P., {\"O}zel, F., 
\& Chakrabarty, D.\ 2002b, \apj, 581, 550

\bibitem[Muno et al.(2004)]{muno2004} Muno, M.~P., Galloway, 
D.~K., \& Chakrabarty, D.\ 2004, \apj, 608, 930 

\bibitem[Narayan 
\& Cooper(2007)]{narayan2007} Narayan, R., \& Cooper, R.~L.\ 2007, \apj, 665, 628 

\bibitem[Piro 
\& Bildsten(2005a)]{piro2005a} Piro, A.~L., \& Bildsten, L.\ 2005a, \apj, 619, 1054 

\bibitem[Piro 
\& Bildsten(2005b)]{piro2005b} Piro, A.~L., \& Bildsten, L.\ 2005b, \apj, 629, 438 

\bibitem[Shara(1982)]{shara1982} Shara, M.~M.\ 1982, \apj, 261, 
649

\bibitem[Spitkovsky et al.(2002)]{spitkovsky2002} Spitkovsky, A., 
Levin, Y., \& Ushomirsky, G.\ 2002, \apj, 566, 1018 

\bibitem[Strohmayer et al.(1996)]{strohmayer1996} Strohmayer, T.~E., 
Zhang, W., Swank, J.~H., et al.\ 1996, \apjl, 469, L9 

\bibitem[Strohmayer \& Markwardt(1999)]{strohmayer1999} Strohmayer, T.~E., \& Markwardt, C.~B.\ 1999, \apjl, 516, L81 

\bibitem[Strohmayer 
\& Bildsten(2006)]{strohmayer2006} Strohmayer, T., \& Bildsten, L.\ 2006, Compact stellar X-ray sources, 113 

\bibitem[Watts et al.(2005)]{watts2005} Watts, A.~L., Strohmayer, 
T.~E., \& Markwardt, C.~B.\ 2005, \apj, 634, 547 

\bibitem[Watts et al.(2008)]{watts2008} Watts, A.~L., Patruno, A., \& van der Klis, M.\ 2008, \apjl, 688, L37

\bibitem[Watts(2012)]{watts2012} Watts, A.~L.\ 2012, \araa, 50, 609 

\bibitem[Watts et al.(2015)]{watts2015} Watts, A., Espinoza, C.~M., Xu, R., et al.\ 2015, Advancing Astrophysics with the Square Kilometre Array (AASKA14), 43 

\bibitem[Wijnands 
\& van der Klis(1998)]{wijnands1998} Wijnands, R., \& van der Klis, M.\ 1998, \nat, 394, 344 

\bibitem[Wijnands et al.(2001a)]{wijnands2001} Wijnands, R., Miller, J.~M., Markwardt, C., Lewin, W.~H.~G., \& van der Klis, M.\ 2001a, \apjl, 560, L159 

\bibitem[Wijnands et al.(2001b)]{rudy2001} Wijnands, R., 
Strohmayer, T., \& Franco, L.~M.\ 2001b, \apjl, 549, L71 

\bibitem[Worpel et al.(2015)]{worpel2015} Worpel, H., Galloway, 
D.~K., \& Price, D.~J.\ 2015, \apj, 801, 60 

\bibitem[Zhang et al.(2013)]{zhang2013} Zhang, G., M{\'e}ndez, 
M., Belloni, T.~M., \& Homan, J.\ 2013, \mnras, 436, 2276 

\end{thebibliography}
\bibliographystyle{apj}

\end{document}